\documentclass[preprint,12pt]{elsarticle} 
\usepackage{amssymb}
\usepackage{wasysym} 
\usepackage{relsize} 
\journal{Astroparticle Physics}
\newcommand{\be}{\begin{equation}}
\newcommand{\ee}{\end{equation}}
\newcommand{\simless}{\lower.5ex\hbox{$\; \buildrel < \over \sim\;$}}
\newcommand{\simgreat}{\lower.5ex\hbox{$\; \buildrel > \over \sim\;$}} 
\newcommand{\mpro}{ m_{\rm p}}

\newcommand{\mesa}{\small{\sl MESA}} 

\newcommand{\metal}{{\cal Z}}

\newcommand{\seff}{{\cal S}_{\rm eff}} 
\newcommand{\teff}{T_{\rm eff}} 
\newcommand{\dineutron}{{^2{n}}}
\newcommand{\diproton}{{^2{p}}}

\newcommand{\deuteron}{n{\kern-0.22em}\ast{\kern-0.22em}p}
\newcommand{\ftime}{\langle ft \rangle} 
\newcommand{\gfactor}{\langle G \rangle} 
\newcommand{\sigpd}{\langle\sigma{v}\rangle_{\diproton{e}}} 
\newcommand{\sigpp}{\langle\sigma{v}\rangle_{pp}} 
\newcommand{\conlum}{ {\cal C}} 
\newcommand{\propro}{$p$-$p$~} 

\newcommand{\gstar}{{g_{\star}}} 
\newcommand{\mpl}{{m_{\rm pl}}} 

\begin{document}

\begin{frontmatter}

\title{{\bf Effects of Bound Diprotons and Enhanced 
Nuclear Reaction Rates on Stellar Evolution}}

\author[1,2]{Fred C. Adams}
\author[3]{Alex R. Howe}  
\author[4]{Evan Grohs}
\author[5]{and George M. Fuller} 

\address[1]{Physics Department, University of Michigan, Ann Arbor, MI 48109} 
\address[2]{Astronomy Department, University of Michigan, Ann Arbor, MI 48109} 
\address[3]{NASA Goddard Space Flight Center, 8800 Greenbelt Rd, 
Greenbelt, MD 20771}
\address[4]{Department of Physics, University of California Berkeley, 
Berkeley, California 94720} 
\address[5]{Physics Department, University of California San Diego, 
La Jolla, CA 92093} 

\begin{abstract}
Deuterium represents the only bound isotope in the universe with
atomic mass number $A=2$. Motivated by the possibility of other
universes, where the strong force could be stronger, this paper
considers the effects of bound diprotons and dineutrons on stars. We
find that the existence of additional stable nuclei with $A=2$ has
relatively modest effects on the universe.  Previous work indicates
that Big Bang Nucleosynthesis (BBN) produces more deuterium, but does
not lead to catastrophic heavy element production. This paper revisits
BBN considerations and confirms that the universe is left with an
ample supply of hydrogen and other light nuclei for typical
cosmological parameters. Using the \mesa~ numerical package, we carry
out stellar evolution calculations for universes with stable
diprotons, with nuclear cross sections enhanced by large factors $X$.
This work focuses on $X=10^{15}-10^{18}$, but explores the wider range
$X$ = $10^{-3}-10^{18}$.  For a given stellar mass, the presence of
stable diprotons leads to somewhat brighter stars, with the radii and
photospheric temperatures roughly comparable to thoese of red
giants. The central temperature decreases from the characteristic
value of $T_c\approx1.5\times10^7$ K for hydrogen burning down to the
value of $T_c\approx10^6$ K characteristic of deuterium burning. The
stellar lifetimes are smaller for a given mass, but with the extended
possible mass range, the smallest stars live for trillions of years,
far longer than the current cosmic age. Finally, the enhanced cross
sections allow for small, partially degenerate objects with mass
$M_\ast=1-10M_J$ to produce significant steady-state luminosity 
and thereby function as stars.
\end{abstract}  


\end{frontmatter}

\newpage 

\section{Introduction} 
\label{sec:intro}  

The laws of physics, as realized in our universe, conspire to produce
a long-lived expanding space-time that contains a wide range of
astrophysical structures, from galaxy clusters down to small, rocky
planets. An ongoing debate in cosmology considers whether our local
universe is one out of many \cite{ellis2004,davies2004,garriga2008,
hall2008,linde2017} and whether or not the laws of physics could 
be different in other (causally disconnected) regions.  A related 
question is the degree to which the laws of physics are fine-tuned 
for the formation of astrophysical structures and ultimately the
development of observers (see \cite{adamsreview} for a recent
review). A common example of possible fine-tuning concerns the bound
state of the diproton (denoted here as $\diproton$).  In our universe,
deuterium is the only bound $A=2$ nucleus. If diprotons were bound and
long-lived, then nuclear reactions could take place through the strong
and electromagnetic interactions, without the weak force. The
resulting cross sections could thus be larger by enormous factors,
$X$, typically estimated as $X\sim10^{15}-10^{18}$.

Many authors (starting with \cite{dyson}) claim that with such large
cross sections, bound diprotons would have catastrophic consequences
for the cosmos. One concern is that during the epoch of Big Bang
Nucleosynthesis (BBN), all of the protons could be processed into
helium, and perhaps heavier elements, leaving behind no hydrogen to
produce water at later epochs. Similarly, in stellar interiors, the
nuclear reaction rates could be so large that stars would have short
lifetimes. Most authors make assertions of this nature by quoting
earlier claims, but do not perform detailed calculations of either BBN
or stellar evolution \cite{davies1972,bartip,pochet,tegmark1998,hogan,
  reessix,dentfair,tegmark2006}, although some treatments are more
nuanced \cite{schellekens,donoghue}. In contrast, recent work using a
semi-analytic model for stellar structure \cite{adams2008,adams2016}
indicates that a wide range of parameter space allows for stable,
long-lived stars in universes with stable diprotons
\cite{barnes2015,adamsreview}. Additional work finds that stable
diprotons will not compromise BBN \cite{bradford,macdonaldmullan}, in
that large mass fractions of hydrogen will remain. The first paper
\cite{bradford} also argues that diprotons will prevent the normal
operations of stars, whereas the second \cite{macdonaldmullan} points
out that diprotons produced during the BBN epoch can subsequently
decay through the weak interaction and thereby produce large amounts
of deuterium (which is still expected to be {\it more} stable).
Similarly, bound dineutrons (denoted as $\dineutron$) have only a 
modest effect on BBN yields \cite{kneller}, leading to a fractional
change in the helium abundance of $\sim10\%$ (see also
\cite{leedineutron}). Notice also that BBN becomes ineffective in
universes where the baryon to photon ration $\eta$ is smaller than
that of our universe \cite{weakless,grohsweakless,howeweakful}. The
finding that BBN is only modestly affected by bound diprotons can be
understood as follows. If the diproton is a bound state, then protons
can interact through the strong force, with a greatly enhanced cross
section.  However, neutrons are present during BBN, so that $n$-$p$
reactions (which also do not require the weak interaction and have no
coulomb barrier) are already dominant.

Given the contradictory claims in the previous literature, the goal of
this paper is to revisit the previous BBN findings and to carry out
detailed numerical simulations of stellar evolution for universes with
stable diprotons. Here we use the publicly-available \mesa~ code
\cite{mesa,mesa2}. This state-of-the-art numerical package is modified
to include the greatly enhanced nuclear reaction cross sections
arising due to stable diprotons and dineutrons.

In this paper, the key assumption is that diprotons are bound, but the
cross sections and binding energies of other nuclei are only modestly
affected. In our universe, dineutrons are unbound by an energy
increment of order 100 keV, whereas diprotons have an additional
Coulomb energy of order 500 keV \cite{davies1972}.  Diprotons can
become bound with an increase in the strong force coupling constant of
order 10\% \cite{bartip,reessix}.  The main implication of bound
diprotons is that nuclear reactions can take place without the weak
force, so that the cross section for the first step of nucleosynthesis
is greatly enhanced.  The increased binding energy of diprotons would
result in comensurate increases in the binding energy of other nuclei,
but these changes are relatively small: In this context, the main
nucleosynthesis product is helium, with binding energy 28.3 MeV, which
could be increased by up to 10\%.  The additional energy available
from the larger binding energy leads to somewhat longer stellar
lifetimes. Since this effect is relatively small, and uncertain, this
paper uses standard values for the binding energies of all nuclei
except the diprotons and dineutrons. As a result, our estimates for
stellar lifetimes should be considered as lower limits, where the true
lifetimes are expected to be longer by a few percent.

This paper is organized as follows. Section \ref{sec:bbn} discusses
the implications of bound diprotons on Big Bang Nucleosynthesis and
finds that the light element abundances are only modestly affected.
The reaction rates for stellar evolution, along with our approach to
incorporating them, are then discussed in Section \ref{sec:stars}.
The numerical results from the \mesa~ simulations are presented in
Section \ref{sec:mesa}, including the modified main sequence, the time
evolution of central temperature and luminosity, the mass-luminosity
relationship, and the effects of varying the level of enhancement for
the cross sections.  The paper concludes in Section \ref{sec:conclude}
with a summary of results and a discussion of their implications. For
completeness, \ref{sec:newnukes} presents the additional reactions
arising for $A=2$ nuclei, with a focus on those relevant for BBN and
stellar nucleosynthesis. Additional Appendices provide further detail
concerning stellar evolution with bound diprotons, including an
explanation for the nearly constant photospheric temperatures seen in
the simulations (\ref{sec:contemp}), a discussion of partially
degenerate planetary mass ``stars'' (\ref{sec:degenuke}), and the
logarithmic dependence of the stellar lifetime on the nuclear
enhancement factor (\ref{sec:timevx}).

\section{Big Bang Nucleosyntheis Considerations}
\label{sec:bbn} 

In this section we consider the ramifications of bound diprotons on
BBN.  If we suppose that a diproton is bound in our nuclear framework,
then we can extend the scenario to include bound dineutrons. For a bound
diproton, we estimate the Coulomb repulsion energy to be 
\be
  V_{\diproton}=\frac{e^2}{4\pi r}\sim 0.36\left(\frac{4\,{\rm fm}}{r}\right)
\,{\rm MeV},
\ee
where $r$ is the mean separation of the two nucleons. For the sake of
definiteness, we consider values $V_{\diproton}\simeq0.5\,{\rm MeV}$, so
that the binding energy of diprotons will be 0.5 MeV less than that of
dineutrons.  In addition, we will require the binding energy of
diprotons to be small enough such that deuterium remains the most tightly
bound (lowest energy) $A=2$ state. For purposes of estimating BBN
effects in this section, we fix the diproton binding energy such that
the beta decay reaction $\diproton\rightarrow d+e^{+}+\nu_e$ is not
energetically possible. As a result, diprotons have to capture leptons
in order to transmute into deuterons.\footnote{This assumption is made 
for simplicity. But even if diprotons can undergo beta decay, the 
lepton (mainly electron) capture reactions are still likely to be the 
dominant channel of transmutation. This also represents a relatively weakly
bound diproton, which would be a smaller difference from our universe.} 
For the diproton to deuteron
reaction to be exothermic, we can relate the binding energy of the
diproton to that of the deuteron, 
\be
  B_{\diproton} < B_{d} - (m_n-m_p) \simeq 0.93\,{\rm MeV},
\ee
where $m_n$ and $m_p$ are the masses of the neutron and proton, 
respectively. To prevent diproton beta decays, we also require
\be
  B_{\diproton} > B_{d} - (m_n-m_p) - m_e \simeq 0.42\,{\rm MeV},
\ee
where $m_e$ is the mass of the electron.  We have ignored the mass of
the neutrino in both of the above inequalities.  With respect to the
above considerations, we take the binding energies of the diproton and
dineutron to be the following: 
\be 
B_{\diproton} = 0.5\,{\rm MeV}
\qquad {\rm and} \qquad 
B_{\dineutron} = 1.0\,{\rm MeV}.
\ee
We preserve the deuteron binding energy at $B_d=2.2\,{\rm MeV}$.

All three dinucleon abundances will be in weak equilibrium with one
another at sufficiently high temperatures. For $\diproton$ and $d$, we
can calculate the abundance ratio with a Saha-like equation for the
following weak interaction 
\be 
e^- + \diproton \longleftrightarrow d + \nu_e 
\qquad \Rightarrow \qquad 
\mu_e + \mu_{\diproton} = \mu_d + \mu_{\nu_e},
\ee
where $\mu_i$ is the chemical potential of the relevant particle
species $i$.  Using Maxwell Boltzmann statistics for the baryonic
states, we can calculate the abundance ratio of diprotons to 
deuterium, 
\be\label{eq:weak_dip}
\frac{Y_{\diproton}}{Y_d} = \frac{g_{\diproton}}{g_d}
\exp\left[\frac{B_{\diproton}-B_d}{T}+
\frac{m_n-m_p}{T}-\phi_e+\xi_{\nu_e}\right].
\ee
There exists a similar expression for the dineutron to deuteron ratio
\be\label{eq:weak_din}
\frac{Y_{\dineutron}}{Y_d} = \frac{g_{\dineutron}}{g_d}
\exp\left[\frac{B_{\dineutron}-B_d}{T}-
\frac{m_n-m_p}{T}+\phi_e-\xi_{\nu_e}\right].
\ee
In equations (\ref{eq:weak_dip}) and (\ref{eq:weak_din}), $g_X$ are
the spin degrees of freedom of nuclide $X$, $\phi_e=\mu_e/T$ is the
electron degeneracy parameter, and $\xi_{\nu_e}=\mu_{\nu_e}/T$ is the
electron-neutrino degeneracy parameter.  In writing both equations, we
have assumed chemical equilibrium for a given lepton and anti-lepton.
As the temperature decreases, the weak-interaction rates which govern
the interconversion between free-neutrons and free-protons will fall
out of equilibrium.  In the standard cosmology, this process begins
above a temperature of 1 MeV and persists until the formation of
$^4{\rm He}$ below a temperature of 100 keV.  In the presence of bound
$\diproton$ and $\dineutron$, the free-neutron-to-free-proton ratio
may maintain equilibrium to lower temperatures if the weak-interaction
rates which govern the interconversion between the dinucleon states
are fast.  The neutron-to-proton interconversion would occur, e.g.,
via the following sequence
\be
  n(p,\gamma)d(e^+,\overline{\nu}_e)\diproton(\gamma,p)p.
\ee
For the above sequence to be an efficient pathway to weak equilibrium,
the weak interaction and radiative capture/photo-dissociation
reactions must all proceed rapidly.  We will make the assumption that
weak interactions such as $\diproton(e^-,\nu_e)d$ are slow on BBN time
scales, but rapid on stellar evolution time scales.  Therefore, the
weak interactions for dinucleons will not maintain equilibrium for
free neutrons and protons.  However, for the small binding energies of
the dinucleons, those abundances maintain Nuclear Statistical
Equilibrium (NSE) down to low temperatures \cite{bbfh}. The abundance 
$Y_X$ of nuclide $X$ with atomic number $Z$ and atomic mass number $A$
is thus given by 
\be\label{eq:NSE}
Y_X = Y_p^Z Y_n^{A-Z} 2^{(A-3)/2}\pi^{3(A-1)/2}g_XA^{3/2}
\left[\frac{n_b}{(Tm_b)^{3/2}}\right]^{A-1}e^{B_X/T},
\ee
where $n_b$ is the baryon number density, $T$ is the plasma
temperature, $m_b$ is the baryon rest mass, and $B_X$ is the binding
energy of nucleus $X$. The factor in square brackets represents the 
inverse of the entropy, so that the high entropy of the universe acts
to keep the abundance $Y_X$ low. The final exponential factor, due to
nuclear binding, acts in the opposite direction. As a result, $Y_X$ is
determined by the competition between these opposing factors.

The abundances $Y_p$ and $Y_n$ of free protons and free neutrons will
not follow weak equilibrium trajectories at low temperature in this
model. As a result, they are free variables to be specified in
equation (\ref{eq:NSE}). The resulting NSE ratios for the dinucleon
abundances are thus given by 
\be
\frac{Y_{^2p}}{Y_d} = 
\frac{Y_p}{Y_n}\frac{g_{^2p}}{g_d}e^{(B_{^2p}-B_d)/T}
\label{eq:nse_dip},
\ee
and
\be
\frac{Y_{^2n}}{Y_d} = 
\frac{Y_n}{Y_p}\frac{g_{^2n}}{g_d}e^{(B_{^2n}-B_d)/T}
\label{eq:nse_din}.
\ee 
The deuteron is a spin-1 system, so that $g_d=3$.  The diproton and
dineutron are both spin-0 systems, so that 
$g_{\diproton}=g_{\dineutron}=1$. Since the dineutron and diproton
binding energies are less than that of the deuteron, the arguments 
in the above exponentials are stictly negative. If 
$B_{\diproton}<B_{\dineutron}$, we would expect a larger abundance of
dineutrons than diprotons.  However, the diproton abundance is
proportional to $Y_p$, instead of $Y_n$ for the dineutron abundance,
and this difference acts to enhance $Y_{\diproton}$ over
$Y_{\dineutron}$.  At early times ($T>10$ MeV), the neutron-to-proton
ratio stays close to unity and the exponential suppression from the
binding energy expression leads to $Y_{\dineutron}\gtrsim
Y_{\diproton}$.  Conversely, at later times ($T<100$ keV), the neutron
abundance becomes negligible, and it is possible for $Y_{\diproton}$
and $Y_{\dineutron}$ to differ greatly.

For $Y_{\diproton}$ and $Y_{\dineutron}$ to maintain NSE abundances,
the rates of creation and destruction of the nuclei must be large. 
The radiative capture reactions $p(p,\gamma)\diproton$ and 
$n(n,\gamma)\dineutron$ provide one channel for production and
photo-dissociation.  These channels are electromagnetic.  The faster
channels would be the following strong interactions with $d$ in the
final state 
\be
\begin{array}{ll}
  ^2p(n,p)d & \quad N_A\langle\sigma v\rangle_{^2pn} \sim
  7\times10^{8}\,{\rm cm}^3/{\rm s},\\
  ^2n(p,n)d & \quad N_A\langle\sigma v\rangle_{^2np} \sim
  7\times10^{8}\,{\rm cm}^3/{\rm s},
\end{array}
\ee
where $N_A$ is Avogadro's number.  For both reactions above, we 
have taken the thermally-averaged cross section from the reaction  
$^3{\rm He}(n,p)t$ to obtain an estimated strength.  Once the reverse
rates for the above two reactions fall below the Hubble expansion
rate, diproton and dineutron synthesis will become inoperative. 
If there is a large abundance of free protons, an individual proton
can capture on $\dineutron$ and destroy the $\dineutron$ abundance. 
For $\diproton$, if the free-neutron abundance is negligible, the main
pathway for $\diproton$ transmutation would have to be a weak reaction
or a photo-dissociation.  Both pathways are temperature sensitive, so
we would expect a freeze-out of the diproton abundance.

Our goal here is to estimate the dinucleon abundances at the point of
departure from secular equilibrium. As a result, we need to determine
the temperature at which the expressions in equations
(\ref{eq:nse_dip}) and (\ref{eq:nse_din}) no longer hold.

\begin{figure}
\includegraphics[width=\columnwidth]{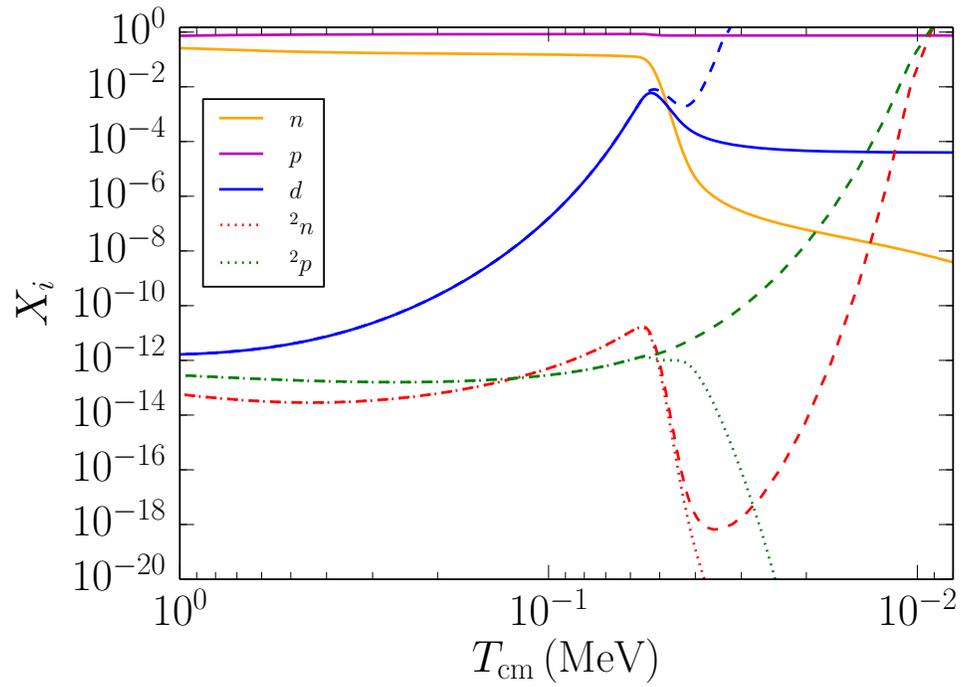}
\vskip-1.50truein
\caption{\label{fig:BBN_nse} Standard BBN abundances (solid lines) as
a function of decreasing $T_{\rm cm}$.  Dashed lines show NSE tracks 
for dinucleons (see equation [\ref{eq:NSE}]) and dotted lines show
secular equilibrium tracks (see equations [\ref{eq:nse_dip}] and
[\ref{eq:nse_din}]).}
\end{figure}

We begin by writing the Hubble expansion rate in terms of temperature 
for radiation-dominated conditions, 
\be
H = \sqrt{\frac{8\pi}{3}\frac{\pi^2}{30}\gstar}\frac{T^2}{\mpl}
\simeq 1.3\,\, {\rm s}^{-1}\left(\frac{\gstar}{43/4}\right)
\left(\frac{T}{1\,{\rm MeV}}\right)^2,
\ee
where $\gstar$ is the statistical degrees of freedom of the radiation
energy density, and $\mpl$ is the Planck mass.  The rate that keeps
$\diproton$ in NSE is
\be
\frac{dY_{\diproton}}{dt} = 
Y_d Y_p n_b\langle\sigma v\rangle_{\diproton n}
e^{-Q_{\diproton n}/T},
\ee
where $Q_{\diproton n}=1.7\,{\rm MeV}$ is the $Q$-value for the
$\diproton(n,p)d$ reaction.  If we use the NSE ratio in equation
(\ref{eq:nse_dip}), we can calculate the time-rate-of-change in the
natural logarithm of $Y_{\diproton}$, 
\be
\frac{d\ln Y_{\diproton}}{dt} 
= Y_n \frac{g_{d}}{g_{\diproton}}e^{(B_d-B_{\diproton})/T}
\frac{2\zeta(3)}{\pi^2}T^3\eta
\langle\sigma v\rangle_{\diproton n}e^{-Q_{\diproton n}/T}
= Y_n \frac{6\zeta(3)}{\pi^2}T^3\eta
\langle\sigma v\rangle_{\diproton n}.
\ee
Once the $\diproton$ production rate falls below the Hubble expansion
rate, $Y_{\diproton}$ no longer maintains NSE. Equating the rate 
$d\ln Y_{\diproton}/dt$ to the Hubble expansion rate, we find 
\be
Y_n \frac{6\zeta(3)}{\pi^2}T^3\eta \langle\sigma v\rangle_{\diproton n}
  = \sqrt{\frac{8\pi}{3}\frac{\pi^2}{30}}\gstar\frac{T^2}{\mpl} \,.
\label{eq:dip_fo}
\ee
This expression can be solved for temperature to obtain 
\be
T \simeq \sqrt{\frac{8\pi}{3}\frac{\pi^2}{30}}
\frac{\pi^2}{6\zeta(3)}\frac{1}{Y_n}
\frac{1}{\eta\langle\sigma v\rangle_{\diproton n}}
\frac{\sqrt{\gstar}}{\mpl} \qquad \qquad \qquad 
\label{eq:t1} 
\ee
$$
\simeq 6\times10^{-9}\,{\rm MeV}\left(\frac{1}{Y_n}\right)
\left(\frac{6\times10^{-10}}{\eta}\right)
\left(\frac{1.16\times10^{-15}\,{\rm cm}^2}
{\langle\sigma v\rangle_{\diproton n}}\right)
\left(\frac{\gstar}{3.36}\right)^{1/2}
$$
$$
\simeq 6\times10^{-2}\,{\rm MeV}\left(\frac{10^{-7}}{Y_n}\right)
\left(\frac{6\times10^{-10}}{\eta}\right)
\left(\frac{1.16\times10^{-15}\,{\rm cm}^2}
{\langle\sigma v\rangle_{\diproton n}}\right)
\left(\frac{\gstar}{3.36}\right)^{1/2}\,.
$$
The right-hand-side of equation (\ref{eq:t1}) includes the neutron
abundance, which is an implicit function of temperature. As a result,
we caution that this expression should be taken only as a guide.  For
example, the first numerical expression is written in terms of the
temperature where $Y_{\diproton}$ freezes-out if $Y_n=1$.  Similarily,
the second numerical expression gives the temperature where
$Y_{\diproton}$ freezes-out if $Y_n=10^{-7}$. Figure \ref{fig:BBN_nse} 
plots the evolution tracks of the various light nuclei mass fractions
as a function of decreasing comoving temperature parameter, 
$T_{\rm cm}$.  The mass fraction is defined using the abundance 
\be
X_i = A_i Y_i,
\ee 
for atomic mass number $A_i$, while $T_{\rm cm}$ is a proxy for the
scale factor and differs from the plasma temperature due to heating of
the plasma from electron-positron annihilation \cite{xmelec}.  The
solid lines are from an actual standard BBN calculation without bound
diprotons or dineutrons \cite{grohs2015}.  The dotted lines give the
mass fractions of diprotons and dineutrons assuming the secular
equilibrium in equations (\ref{eq:nse_dip}) and (\ref{eq:nse_din}).
In addition, we plot the NSE mass fractions for the dinucleon states
from equation (\ref{eq:NSE}).  By inspection of Figure
\ref{fig:BBN_nse}, we see that the value for the $\diproton$
freeze-out temperature for $Y_n$ = $10^{-7}$ is an overestimate, while
the value for $Y_n=1$ is a gross underestimate (using numerical values
from equation [\ref{eq:t1}]).  However, if the diproton abundance
follows the dotted curve from secular equilibrium, we see that the
addition to the primordial deuterium abundance from diproton
transmutation is small and of little consequence to the initial
composition of stars at later epochs.

The dineutron abundance would follow a similar expression to the
diproton abundance in equation (\ref{eq:dip_fo}), except for the
replacement of the free-neutron abundance with the free-proton
abundance. The proton abundance $Y_p$ also has time dependence, but
the mass fraction always stays close to unity in standard BBN as seen
in Figure \ref{fig:BBN_nse}.  As equation (\ref{eq:t1}) indicates a
low temperature for diproton freeze-out if $Y_n=1$, we conjecture that
the dineutron abundance would stay in secular equilibrium to very low
temperatures.  Therefore, we expect an insignificant addition to
primordial deuterium from dineutron transmutation, as indicated by 
the dotted curves in Figure \ref{fig:BBN_nse}. 

Our arguments in this section rely on a number of assumptions.  First,
we have used NSE and secular equilibria between the $A=2$ mass states
to evolve the diproton and dineutron abundances.  Figure
\ref{fig:BBN_nse} shows that all three of the $A=2$ mass fractions
align with the NSE values down to a low temperature $T\lesssim100$ 
keV.  However, immediately after the point of departure from NSE,
the NSE values and actual/secular equilibrium values differ greatly.
Second, both the NSE and secular equilibrium expressions are
exponentially sensitive to binding energy.  Picking different values
of the diproton and dineutron binding energy will change the point of
departure from NSE.  Third, we used temperature-independent cross
sections based on strong interactions for the
$\diproton\leftrightarrow d\leftrightarrow\dineutron$ reactions.
Introducing temperature dependence or changing the overall strength of
the cross sections will also influence $A=2$ nucleosynthesis.
Finally, we have only considered the reactions that create diprotons
and dineutrons out of free particles or transmute them to deuterium.
We have not considered other strong reactions, such as
$d(\dineutron,n)t$, which could lead to an earlier epoch of
$^4{\rm He}$ synthesis and hence a larger primordial abundance.

Addressing the above criticisms requires a numerical calculation
beyond our equilibrium estimates, as well as an extensive exploration
of parameter space.  Nevertheless, for the nuclear binding energies
considered here, we estimate that the primoridal abundance of free
protons, deuterium, and helium will be approximately the same as from
standard BBN in our universe. The high entropy of the universe 
prevents the conversion of most of the protons into bound states, 
as well as the contruction of larger nuclei. The results from BBN 
could thus be different for universes in which the entropy is 
substantially lower. 

\section{Stellar Nuclear Reactions} 
\label{sec:stars} 

This section considers the nuclear reactions that are relevant for
stellar interiors. Only the reactions involving diprotons are
significant. As shown above, BBN generally does not produce enough
dineutrons to affect stellar evolution, and any dineutrons that are
produced will have decayed into deuterium. The remaining reactions
that affect the production of helium are discussed below.

The process starts with the original production reaction, 
\be 
p + p \longrightarrow \diproton + \gamma \,. 
\label{diproduction} 
\ee 
The weak decay of the diproton involves the production of a positron,
which requires a minimum amount of energy (0.511 MeV).  On the other
hand, the center of the star contains a population of free electrons,
so we expect the tranformation reaction that turns diprotons into
deuterium to be given by electron capture, 
\be
e^{-} + \diproton \longrightarrow d + \nu_e \,. 
\label{dipcapture} 
\ee
Although this reaction involves the weak force, it does not require
the particles to overcome a Coulomb barrier, so the rate is much
larger than the corresponding reaction at the start of the \propro
chain. Because the charges lead to an attractive force, the cross
section is increased further due to the Sommerfeld enhancement. The
remaining new reaction involves the interaction of the diproton with
the deuteron to produce $^3$He, 
\be
\diproton + d \longrightarrow ~^3{\rm He} + p \,.
\label{dipdeut} 
\ee
However, this new reaction is likely to be slower than the 
standard interaction of the deuteron, i.e., 
\be
d + p \longrightarrow ~^3{\rm He} + \gamma \,.
\label{prodeut} 
\ee
In the Sun, for example, the deuteron has a typical lifetime of about
4 seconds, which is much shorter than the typical interaction time of
the original protons (a few Gyr). Moreover, the usual deuterium
burning reaction (\ref{prodeut}) is likely to dominate over equation
(\ref{dipdeut}) because of the increased Coulomb repulsion in the
latter. The resulting $^3$He then interacts through the usual channel, 
\be
~^3{\rm He} + ~^3{\rm He} \longrightarrow  
~^4{\rm He} + 2p + \gamma \,. 
\label{he334} 
\ee
This set of reactions corresponds to the so-called PPI chain in the
Sun, where this channel accounts for about 85\% of the helium
production \cite{clayton,kippenhahn,phillips,hansen}. For the cases of
interest, we find that the operating temperature in stellar cores is
lower (see Section \ref{sec:mesa}) so that the suppression due to
Coulomb barrier penetration is more severe, and the PPI chain should
dominate over other channels by an even larger margin than in the 
Sun.\footnote{Recall that the CNO cycle is more temperature sensitive
than the \propro chain. The lower central temperatures for stars with 
diprotons thus results in a suppression of CNO reactions. } 
For completeness, we note that the presence of stable diprotons allows
for the alternate reaction $~^3$He($^3$He,$\diproton$)$^4$He, which
produces a diproton instead of two separate protons.  In our universe
the original reaction (\ref{he334}) has a yield of 12.86 MeV, which is
much larger than the binding energy of $\diproton$, so that diproton
production is expected to be minimal.

Given the above reactions, nuclear burning in stars containing stable
diprotons could in principle proceed in different ways:

[$\mathbb{I}$] In the first scenario, the production reaction
(\ref{diproduction}) takes place, but the subsequent weak interaction
(\ref{dipcapture}) is slow.  In this case, the star burns its protons
into diprotons, but the diproton products remain inert (much like
helium collects in stellar cores in our universe, but does not react
until later stages of stellar evolution).  The energy yield for this
scenario is determined by the binding energy of the diproton.  This
binding energy is expected to be comparable to that of deuterium in
our universe ($B_d\approx2.2$ MeV), about $\sim1$ MeV per particle. 

[$\mathbb{II}$] In the second scenario, the weak interaction
(\ref{dipcapture}) proceeds rapidly enough that diprotons are
converted into deuterium as they are produced. The subsequent reaction
(\ref{prodeut}) then produces $^3$He, in close analogy to the
processes taking place in stellar cores in our universe, albeit with
greatly enhanced cross sections. However, the temperature is generally
too low for the $^3$He to be processed into $^4$He through equation
(\ref{he334}). The stellar core is thus converted into $^3$He, but it
burns in a later evolutionary stage.  The binding energy of $^3$He is
about 7.7 MeV in our universe, and is expected to be somewhat larger
in a universe with stable diprotons. The energetic yield for this
scenario is $\sim3$ MeV per particle.

[$\mathbb{III}$] For some range of parameters, the nuclear burning
temperature for the processes of equations (\ref{diproduction} --
\ref{prodeut}) could be high enough that $^3$He is processed as it is
synthesized. In this case, the original protons are transformed into
$^4$He in a single stage of stellar nucleosynthesis, roughly analogous
to the \propro chain in the Sun. The energetic yield for this scenario
is thus $\sim7$ MeV per particle (although the intermediate steps are
different than for the standard \propro chain).

Which of the scenarios is in play is determined by the relative 
rates of diproton destruction and production, where this 
ratio is given by 
\be
{\cal R} = {n_e n_{\diproton} \sigpd \over n_p^2 \sigpp} \approx
{n_{\diproton} \sigpd \over n_p \sigpp} \,.
\ee
The following subsections estimate the diproton production cross
section (denominator, Section \ref{sec:dicross}) and the diproton
destruction cross section (numerator, Section \ref{sec:destruct}).
These considerations indicate that the diproton destruction rate (via
deuterium production) is fast enough to keep pace with diproton
production, so that we expect $^3$He to be produced promptly through
reactions (\ref{diproduction}), (\ref{dipcapture}), and
(\ref{prodeut}). On the other hand, the stellar evolution simulations 
of the following section indicate that the nuclear burning temperature 
is only $T_c\sim10^6$ K with the expected enhancement of the cross 
sections. This temperature is too low for $^3$He to react, so that 
$^4$He production takes place at a later epoch. As a result, stars 
are expected to follow scenario $\mathbb{II}$.


\subsection{Cross Section for Diproton Production} 
\label{sec:dicross} 

In general, the velocity averaged cross section for stellar 
nuclear reactions can be written in the from 
\be
\langle \sigma v \rangle = 
\left({8\over\pi m_R}\right)^{1/2} (kT)^{-3/2} 
\int\sigma E \exp[-kT/E] dE \,,
\ee
where $m_R$ is the reduced mass of the two interacting particles
\cite{clayton,phillips}.  For charged particles undergoing
non-resonant interactions, the cross section can be written as 
\be
\sigma(E) = {S(E) \over E} \exp\left[ - (E_G/E)^{1/2} \right] \,,
\ee
where $E_G$ is the Gamow energy $E_G=2(\pi\alpha Z_1 Z_2)^2m_R\,c^2$.
After integrating over the thermal distribution of particle energies,
the resulting cross section (weighted by velocity) can be written in 
the form 
\be
\langle \sigma v \rangle = {8 \over 9} 
\left({2\over 3 E_G m_R}\right)^{1/2} 
\seff \Phi^2 \exp[-\Phi] \,. 
\ee
where 
\be
\Phi \equiv 3(E_G/4kT)^{1/3} \,.
\label{phidef} 
\ee

For $p(p,e^{+}\nu)d$, $m_R=\mpro/2$ = 469 MeV and $E_G$ = 493 keV. If
the reaction produces diprotons instead, the reduced mass and Gamow
energy remain the same.  The coefficient $\seff$ is generally
expressed in units of MeV-barn, where values for reactions of interest
are given in Table 1. With this set of units of $\seff$, the cross
section can be written in the form  
\be
\langle \sigma v \rangle = 
(1.43 \times 10^{-15} {\rm cm}^3 {\rm s}^{-1}) 
\seff \Phi^2 \exp[-\Phi] \,. 
\ee
Finally, we can write the parameter $\Phi$ in the form 
\be
\Phi = \lambda \Phi_0 \qquad {\rm where} \qquad 
\lambda = (Z_1 Z_2)^{2/3} A^{1/3} \,, 
\ee
where $A=A_1A_2/(A_1+A_2)$. The value of $\Phi_0$ is 
the same for all reactions (at a given temperature), i.e.,
\be
\Phi_0 \approx 19.76 \left({T \over 10^7\,{\rm K}}\right)^{-1/3}\,. 
\ee 
The expression for the cross section is then 
\be
\langle \sigma v \rangle = 
(1.43 \times 10^{-15} {\rm cm}^3 {\rm s}^{-1}) 
\seff \lambda^2 \Phi_0^2 \exp[-\lambda\Phi_0] \,. 
\label{crosseval} 
\ee

In our universe, the cross section for the reaction $p(d,\gamma)^3$He
is larger than that for $p(p,e^{+}\nu)d$ by a factor
$X\approx7\times10^{17}$ (see Table 1). As a first approximation, we
expect a comparable enhancement factor for the reaction
(\ref{diproduction}) that produces diprotons. However, the binding
energy of $^3$He is (most likely) larger than that of the diproton,
and nuclear reaction rates can vary by a few orders of magnitude even
for analog reactions.  For example, the cross section for
$t{(p,\gamma)}^4$He is an order of magnitude smaller than that for the
less energetic reaction $d{(p,\gamma)}^3$He.  The
$p(p,\gamma)\diproton$ reaction will be less energetic still, so we
expect the cross section and enhancement factor to be smaller than the
above estimate.  For the sake of definiteness, this paper focuses on
the values $X=10^{15}$ and $X=10^{18}$ for the stellar evolution
calculations of the following section. Nonetheless, a wider range of
enhancement factors is possible. It is useful to see how stellar
evolution changes over a larger parameter space, so we also (briefly)
consider stars with $X=10^{-3}-10^{18}$ (see Section
\ref{sec:xvariation}). 

For reactions that process protons into diprotons, the nuclear burning
temperature is about $T\sim10^6$ K, so that $\Phi_0\approx40$. If we
adopt enhancement factors in the range $X=10^{15}-10^{18}$, which 
corresponds to $\seff\approx3\times10^{-10}-3\times10^{-7}$ MeV-barn, 
the resulting cross section from equation (\ref{crosseval}) can be 
estimated as  
\be
\sigpp \sim 3 \times (10^{-36}-10^{-39}) {\rm cm}^3 {\rm s}^{-1} \,.
\label{sigppnumber} 
\ee 

\bigskip 
\begin{table}  
\centering 
\medskip 
\begin{tabular}{llll} 
\\
\hline 
\hline 
reaction & $\seff$ (MeV-barn) & $\lambda$ & $X$ \\ 
\hline 
$pp\to~{d}$ & $3.36\times10^{-25}$ & 1 & 1 \\ 
$pd\to~^3{\rm He}$ & $2.50\times10^{-7}$ & 1.26 & $7\times10^{17}$\\ 
$pp\to~\diproton$ & $3\times(10^{-10}-10^{-7})$ & 1 & $10^{15}-10^{18}$\\ 
\hline 
\hline 
\end{tabular}
\caption{\bf Cross Section Parameters }
\label{table:crosses} 
\end{table}  

\subsection{Weak Interaction Cross Section for Diproton Destruction} 
\label{sec:destruct} 

Weak interactions for stellar nucleosynthesis have been studied
previously \citep{fullerfowler,fullerfowlertwo}. Following this
prior work, we define the quantity $\ftime$ such that the decay
rate is given by 
\be
\Gamma = 1/\tau = {\ln 2 \over \ftime} f \,,
\ee
where $f$ is the phase space density factor. In approximate terms, the
quantity $\ftime$ plays the role of the half-life divided by the phase
space function.  The matrix element for the interaction is given by 
\be
|M_{GT}|^2 = \sum_{if} {n_p^i n_n^f \over 2j_f+1} 
|M_{GT}^{sp}|_{if}^2 \,,
\ee
where the matrix element and the $\ftime$ factor are related 
through the approximate expression 
\be
\log_{10} \ftime = 3.596 - \log_{10} |M_{GT}|^2 \,.
\ee
After considerable analysis \citep{fullerfowler,fullerfowlertwo}, 
the cross section can be written in the form  
\be
\sigma = 2\pi^2\,(\ln 2)\,{\gfactor \over \ftime}\,
{(E + Q)^2 \over m_e^5} \,,
\label{weakcross} 
\ee
where $\gfactor$ represents an integral over the phase space and is
temperature dependent. Notice that $\gfactor$ is expected to be larger
than unity due to the Sommerfeld enhancement \citep{sommerfeld}.  In
this context, the operating temperature of the star ($T_c\sim10^6$ K)
is low compared to the energy scales of interest ($E\sim1$ MeV
$\sim10^{10}$ K), so that the expression can be evaluated in the low
temperature limit.  The quantity $Q$ is the mass difference between
the initial and final states. Here we are considering the capture
process $\diproton(e^{-},\nu_e)d$, so that the initial mass is given
by $2 m_p - B_{\diproton} + m_e$, whereas the final mass is given by
$m_p + m_n - B_d$.  The difference determines the $Q$ factor, which
becomes 
\be
Q = m_p + m_e - m_n + B_d - B_{\diproton} 
  \approx -0.8 {\rm MeV} + B_d - B_{\diproton}\,.
\ee
In order for the reaction to be enegetically favored, the binding
energy of deuterium must exceed the binding energy of the diproton by
more than 0.8 MeV. In our universe, the $B_d$ exceeds $B_{\diproton}$
by more than 2.2 MeV, so if the binding energies increase together,
diprotons can be transmuted into deuterons through reaction
(\ref{dipcapture}). For the particular choice $B_{\diproton}$ = 0.5 MeV 
(from Section \ref{sec:bbn}), the factor becomes $Q$ = 0.9 MeV. 

The time scale $\ftime/\gfactor$ is determined by the weak interaction.  
As a working approximation, we expect this time scale to be comparable
to the time $\tau_n$ required for neutrons to decay. This latter time 
scale can be written in the form
\be
\tau_n^{-1} = G_F^2 m_e^5 \mathcal{C}_1\,. 
\ee
The dimensionless factor $\mathcal{C}_1$ can be calculated from an
effective field theory of the weak interaction where the quark degrees
of freedom of the proton/neutron have been integrated out.  In our
universe, $\mathcal{C}_1$ = $\lambda_0 (1+3g_A^2)/2\pi^3$, where the
numerical factor $\lambda_0\approx1.636$, and $g_A\approx1.26$ is the
axial-vector coupling factor for nucleons \cite{kolbturn}. With this
approximation, the cross section from equation (\ref{weakcross}) can
be written in the form
\be
\sigma = 2\pi^2 (\ln2) \mathcal{C}_2 G_F^2 (E + Q)^2 
\approx 2 G_F^2 (E + Q)^2 \sim 4 \times 10^{-43} {\rm cm}^2 \,,
\ee
where the new dimensionless factor $\mathcal{C}_2$ includes the
nuclear physics relevant in the alternate universe where the strong
force is stronger and diprotons are bound.\footnote{In general, if the  
strong force is altered, then the weak force is expected to change as
well. Moreover, the axial-vector coupling $g_A$ in our universe
depends on the properties of nucleons, which can also change as the
strong force is varied. However, we expect such changes to be
modest. } For the sake of definiteness, we have used
$\mathcal{C}_1=\mathcal{C}_2$, along with $E+Q$ = 2 MeV, to evaluate
the cross section.  The electron speed is given approximately by 
\be
v_e \approx \left( {3kT \over m_e} \right)^{1/2} \approx 
7 \times 10^8 {\rm cm}\,\,{\rm s}^{-1} \,, 
\ee
so that we have 
\be
\sigpd \sim 3 \times 10^{-34} {\rm cm}^3 {\rm s}^{-1} \,.
\ee
For the expected operating temperature of the star ($T_c\sim10^6$ K),
this (velocity weighted) cross section is larger than that for
producing diprotons by a factor of $10^2-10^5$ over the expected range
of nuclear enhancement factors $X$ (see equation [\ref{sigppnumber}]).
As a result, we expect diprotons to be converted into deuterium as
fast as they are produced. As a working approximation, we can assume
that the cross section for diprotons production corresponds to that
for deuterium production. Moreover, as long as the $e^{-}$ capture
process is rapid compared to diproton production, the exact capture
rate does not matter (the diprotons immediately capture electrons and
become deuterium).  Thus, the net result for stellar nucleosynthesis with
stable diprotons is to make the factor $\seff$ much larger than
in our universe, where we consider enhancement factors in the range
$X=10^{15}-10^{18}$.

The approximations outlined above should remain valid unless the weak
interaction becomes significantly less effective than in our universe.
However, if the rate at which electron capture converts diprotons into
deuterium is too slow, then stars would process protons into
diprotons, but the latter nuclei would remain inert. This alternate
scenario is unlikely.  Specifically, the dimensionless factor
$\mathcal{C}_2$ would have be more than 100 times smaller than the
value ($\mathcal{C}_1$) in our universe, although it is expected to be
{\it larger} due to Sommerfeld enhancement. To study this issue,
future work should carry out a full Hartree-Fock type of calculation,
which is beyond the scope of this present paper. Of course, even if
diprotons cannot be rapidly converted into deuterium, stars would
still operate, albeit with a lower yield.  In this case, the energy
produced per particle would be determined by the binding energy of the
diproton instead of the (larger) binding energy of $^3$He (which holds
when $d+p\to$ $^3$He is rapid, as considered in this paper).  This
difference is a factor of $\sim3$ less energy yield per particle and
would result in a commensurate decrease in stellar lifetimes.

\section{Results from Stellar Evolution Simulations} 
\label{sec:mesa} 

In this section, we present results obtained using the \mesa~ 
stellar evolution code, a state of the art numerical package that 
is publically available \cite{mesa}. Here we focus on scenario
$\mathbb{II}$, where electron capture is fast enough to convert
diprotons into deuterium, but the $^3$He nuclei are not processed
immediately.

To account for this scenario, we enhance the nuclear reaction rates
for the \propro chain by a constant factor $X$. In the usual \propro
chain, the first step is the production of deuterium through the
fusion of two protons. This reaction is slow, due to the necessity of
the weak interaction. In stars with stable diprotons, the protons can
fuse directly into diprotons through the strong interaction. The
resulting diproton will then capture an electron and become deuterium
as before (see equation [\ref{dipcapture}]).  Although this latter
reaction still requires the weak interaction, it is much faster than
usual \propro reactions due to the absence of the coloumb barrier. In
both scenarios, deuterium interacts rapidly with additional protons to
become $^3$He. Since the latter interaction is fast, stellar evolution
codes generally model the reaction as three protons producing
$^3$He. The subsequent reactions involving the conversion of $^3$He
to make $^4$He are then modeled explicitly. The net effect of having
stable diprotons is thus to enhance the reaction rate for $^3$He
production. The results of the previous section indicate that the
enhancement factor is expected to be $X\approx10^{15} - 10^{18}$. 
Here we focus on this range for $X$, but also explore a much wider
range of values.

\subsection{The Main Sequence} 
\label{sec:mainsequence} 

\begin{figure}[tbp]
\centering 
\includegraphics[width=1.0\textwidth,trim=0 150 0 150]{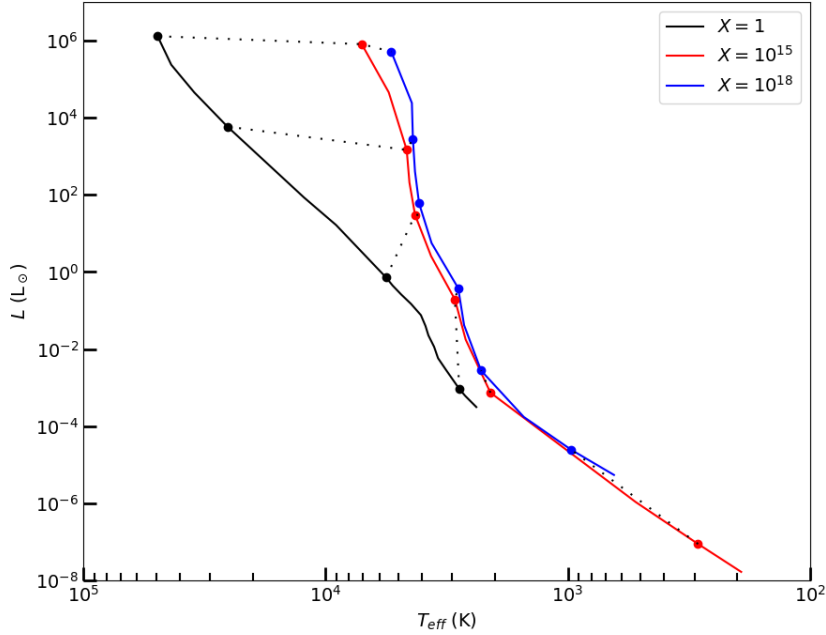}
\vskip 1.0truein 
\caption{Main sequence for stellar nucleosynthesis in our universe 
(left black curve) and for universes with stable diprotons and  
enhanced nuclear reaction rates with $X=10^{15}$ (right red curve) and
$X=10^{18}$ (right blue curve).  The symbols denote specific stellar
mass values: $M_\ast=100M_\odot$ (top), $10M_\odot$, $1M_\odot$,
$0.1M_\odot$ (bottom of black curve), $10^{-2}M_\odot$, and
$10^{-3}M_\odot=1M_J$ (bottom of red curve).  The standard main
sequence extends down to $M_\ast=0.08M_\odot$. The diproton main
sequence contains two distinct branches (see text). For the upper
branch with masses $M_\ast=0.01-100M_\odot$, the main sequence is
steep, with photospheric temperatures spanning a narrow range. For the
lower branch with masses $M_\ast<0.01M_\odot=10M_J$, the stars have 
nearly constant radius, and the main sequence becomes less steep. }  
\label{fig:mainsequence} 
\end{figure}

For \mesa~ simulations with nuclear enhancement factors $X=10^{15}$
and $X=10^{18}$, Figure \ref{fig:mainsequence} shows the main sequence
for stars with stable diprotons compared with those in our
universe. For the standard main sequence in our universe, the stellar
mass range is taken to be $M_\ast=0.08-100M_\odot$.  The upper mass
cutoff is determined by stability considerations. For stars larger
than about $M_\ast\simgreat100$ $M_\odot$, the pressure contribution
due to radiation becomes larger than that due to thermal gas pressure, 
and the star becomes unstable \cite{clayton,kippenhahn,phillips}. 
This upper mass limit is independent of the source of the stellar
luminosity and is expected to be the same for stars with stable
diprotons. In addition, the mass-luminosity relation for sufficiently
massive stars approaches the form $L_\ast\sim4\pi{Gc}M_\ast/\kappa$, 
where $\kappa$ is the opacity. As a result, the brightest stars (with 
$M_\ast=100M_\odot$) have luminosity $L_\ast\sim10^6L_\odot$, as 
shown in Figure \ref{fig:mainsequence}. 

The lower mass cutoff is determined by the minimum mass capable of
producing sustained nuclear fusion. This mass scale depends on the
nuclear burning temperature, which in turn depends on the nuclear
processes that power the star. For hydrogen burning stars in our
universe, the lower mass cutoff is approximately $M_\ast$ = 0.08
$M_\odot$. For stars with with stable diprotons and enhanced reaction
cross sections, the nuclear burning temperature is lower, comparable
to the deuterium burning temperature in our universe ($T_c\sim10^6$
K). With this value for the nuclear burning temperature, the minimum
mass decreases down to $M_\ast$ = 0.01 $M_\odot$ = 10 $M_J$. As shown
in Figure \ref{fig:mainsequence}, the main sequence extends down to
this mass scale. In addition, however, the numerical simulations show
that nuclear burning can be sustained at even lower masses. 

The resulting main sequence for stars with stable diprotons contains two
distinct branches (see Figure \ref{fig:mainsequence}).  The upper
branch corresponds to stars with masses in the range
$M_\ast=0.01-100M_\odot$, and displays a narrow range of photospheric
temperatures. The lower branch includes stars with masses in the range
$M_\ast=0.5-10M_J$, and displays a wider range of (much cooler)
temperatures. These low-mass objects are primarily supported by
electron degeneracy pressure and can thus be considered as the
``degenerate branch'' of the main sequence. In contrast, stellar objects
on the upper main sequence are supported by thermal pressure, like
ordinary stars in our universe.

More specifically, the upper branch of the main sequence is roughly
similar to the main sequence for deuterium burning in our universe.
This result is expected, as the large enhancement factors $X$ for
diproton reactions are motivated by the reaction cross sections for
deuterium \cite{grohsweakless,howeweakful}. The mass range for the
upper main sequence is also consistent with that expected for
deuterium burning.  Moreover, the upper main sequence has a slowly
varying photospheric temperature, nearly independent of stellar mass,
with typical values in the range $T_\ast\approx3000-5000$ K. These
photospheric temperatures not only span a narrower range (even though
the mass range is larger for these diproton stars), but they are also
cooler than ordinary hydrogen-burning stars. The slowly varying
surface temperature leads to a different mass-luminosity relation,
compared with stars in our universe (see Section \ref{sec:masslum} and
\ref{sec:contemp}). The luminosity range for the upper main sequence
is comparable to that of ordinary stars, and spans essentially the
same nine orders of magnitude ($L_\ast=10^{-3}-10^6L_\odot$).

Figure \ref{fig:mainsequence} shows that the upper main sequence 
for both $X=10^{15}$ and $X=10^{18}$ are remarkably similar. Some
uniformity is expected, because stellar properties depend only
logarithmically on the enhancement factor $X$ (\ref{sec:timevx}), but
these main sequences are nearly identical. This result arises from the
nearly constant surface temperatures for these stellar
configurations. Because of the extreme temperature sensitivity of the
opacity, photospheric temperatures cannot fall below a minimum value
\cite{hayashi}, so that the main sequence cannot move farther to the
right in the diagram (for further discussion, see also Section
\ref{sec:vlm} and \ref{sec:degenuke}).

The lower branch of the main sequence shown in Figure
\ref{fig:mainsequence} includes much smaller stars with masses
comparable to Jupiter, i.e., $M_\ast=0.5-10M_J$. This collection of
stellar objects spans an additional four orders of magnitude in
luminosity and displays a wider range of photospheric temperature in
spite of the smaller mass range. The minimum stellar mass for this
lower branch is thus substantially smaller than the minimum mass for
deuterium burning, i.e., $M_\ast\lesssim0.5M_J$. In this regime, the
differences between stars with $X=10^{15}$ and $X=10^{18}$ are more
pronounced, with the luminosities increasingly significantly with
increasing enhancement factor (especially at the lowest stellar masses
$M_\ast\sim1M_J$).

The two parts of the main sequence correspond to different types of
stellar configurations. In the range $M_\ast=0.01-100M_\odot$, the
stars are supported primarily by thermal pressure, where the nuclear
reactions provide heating for the gas that provides the pressure.
These stars are thus like those in our universe.  For the lower
branch, with masses $M_\ast<0.01M_\odot$, the main sequence slope
corresponds to stellar configurations with nearly constant radius.
This behavior indicates that the stars are supported primarily --- but
not entirely --- by degeneracy pressure. These unusual objects are
much like brown dwarfs in our universe, in that hydrostatic
equilibrium does not rely on nuclear reactions, but here the greatly
enhanced nuclear cross sections allow for substantial power generation
(see Section \ref{sec:vlm} and
\ref{sec:degenuke}). For both branches of the main
sequence, stars are able to achieve sustained nuclear power
generation over time scales long enough to be of interest for
cosmology and biology.

\begin{figure}[tbp]
\centering 
\includegraphics[width=1.0\textwidth,trim=0 150 0 150]{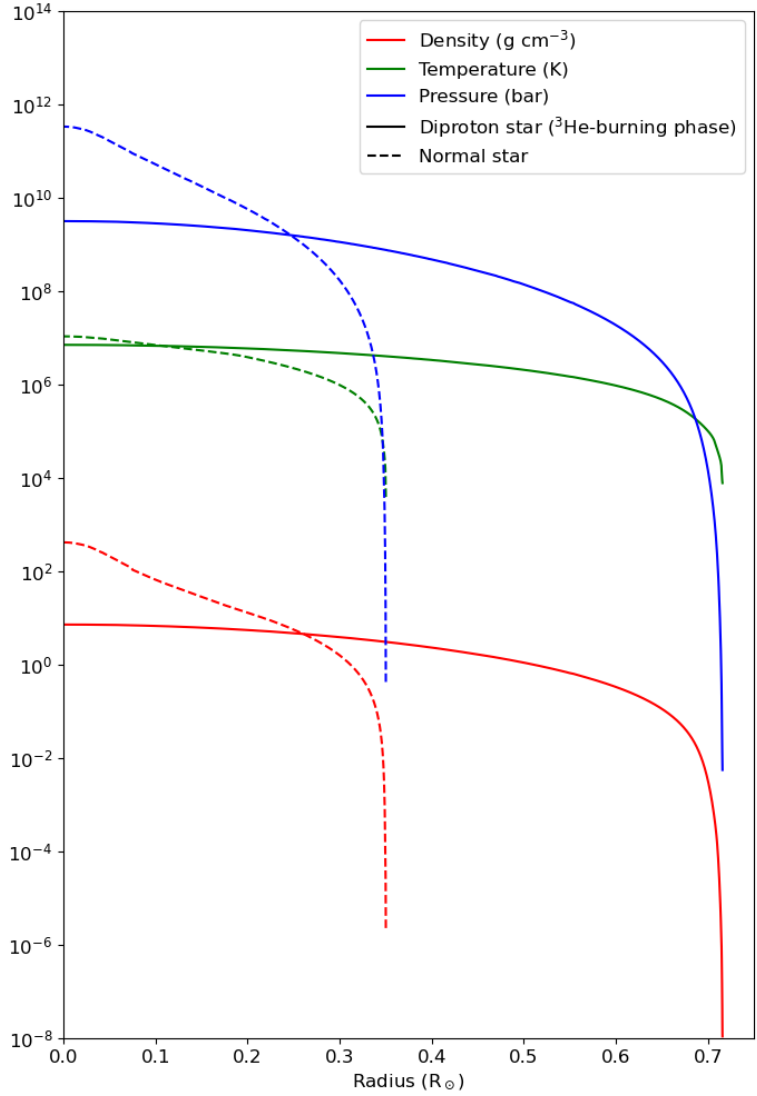}
\vskip 1.0truein 
\caption{Comparison of stellar structure for stars with mass $M_\ast$ 
= 0.3 $M_\odot$ with standard nuclear reactions (dashed curves) and
enhanced rates ($X=10^{15}$, solid curves). The three sets of curves
show the profiles for pressure (upper blue cuves), temperature (middle
green curves), and density (lower red curves) as a function of radial
coordinate within the star. }
\label{fig:starcompare} 
\end{figure}

Figure \ref{fig:starcompare} shows the internal structure of a star
with mass $M_\ast=0.3M_\odot$ for enhanced nuclear reaction rates
(solid curves) and for standard rates (dashed curves). This stellar
mass was chosen because it represents one of the most common stars in
our universe; moreover, its luminosity and lifespan in the alternate
universe are roughly comparable to those of the Sun. The figure shows
the profiles for pressure (blue curves), temperature (green curves),
and density (red curves) for the two stars.  For all three quantities,
the central values are somewhat larger for the star with standard
nuclear reactions rates and the profiles fall off more steeply. The
star with enhanced reaction rates is about twice as large in radius
and displays flatter profiles. The larger radius results in the cooler
surface temperatures (redder colors) shown in the H-R diagram of
Figure \ref{fig:mainsequence}.

\subsection{Time Evolution of Stellar Properties} 
\label{sec:tevolve} 

\begin{figure}[tbp]
\centering 
\includegraphics[width=1.0\textwidth,trim=0 150 0 150]{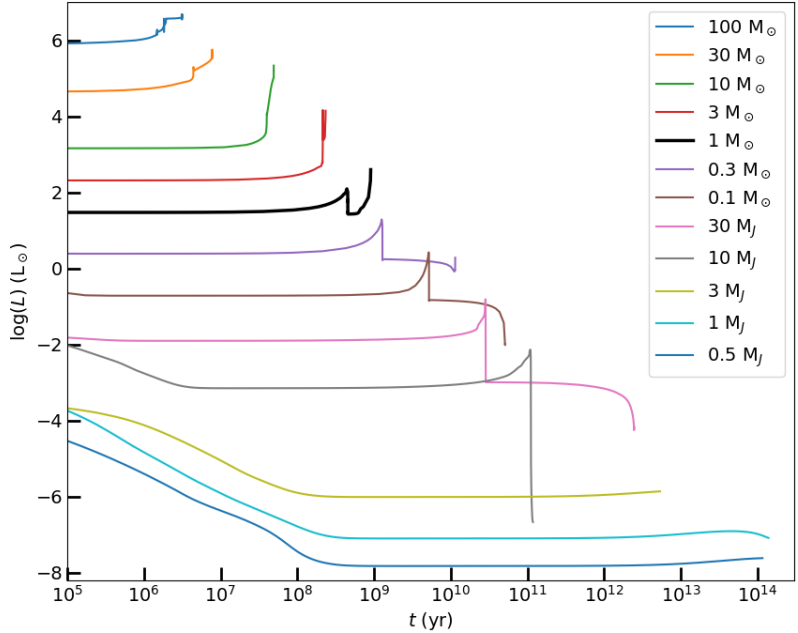}
\vskip 1.0truein 
\caption{Stellar luminosity as a function of stellar age for stars 
with enhanced nuclear reaction rates $X$ = $10^{15}$. } 
\label{fig:lumvtime} 
\end{figure}

\begin{figure}[tbp]
\centering 
\vskip 1.0truein 
\includegraphics[width=1.0\textwidth,trim=0 150 0 150]{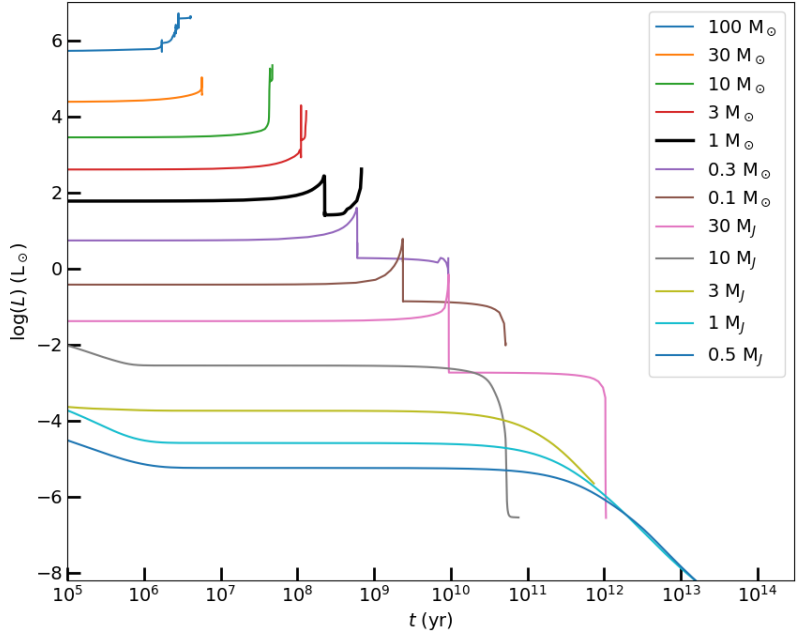}
\vskip 1.0truein 
\caption{Stellar luminosity as a function of stellar age for stars 
with enhanced nuclear reaction rates $X$ = $10^{18}$. } 
\label{fig:lumvtime18} 
\end{figure}

The stellar luminosity is shown as a function of stellar age in
Figures \ref{fig:lumvtime} and \ref{fig:lumvtime18}, for enhancement
factors $X=10^{15}$ and $X=10^{18}$, respectively.  All of the stars
experience a short initial phase of declining brightness corresponding
to pre-main-sequence contraction. This phase is relatively short, so
that only the smallest stars show this behavior over the time scales
shown in the figures. After this early transient phase, all of the
stars experience an extended phase of nearly constant luminosity,
corresponding to main-sequence nuclear reactions (protons $\to$
diprotons $\to$ $^3$He).

For the lower mass stars, the central temperatures for burning protons
into diprotons and then $^3$He, are relatively low, so that the latter
species is not immediately processed into $^4$He.  Instead, the lower
mass stars display two distinct phases, first producing $^3$He as the
product and then later transforming the $^3$He into $^4$He. These
separate phases are evident in both Figures \ref{fig:lumvtime} and
\ref{fig:lumvtime18} for stars with masses in the range $M_\ast$ =
0.03 -- 1 $M_\odot$. For stars of even lower mass (specifically less
than about $M_\ast=0.01M_\odot$ = 10 $M_J$), the central temperature
cannot become high enough to burn $^3$He. These lowest-mass stars only
experience the first of the two phases, and they end their lives with a
$^3$He core.  Higher mass stars with $M_\ast>1M_\odot$ also have
these two phases, but they are much shorter and more blended together
as shown in the figures.

For the highest mass stars, the stellar lifetimes are comparable to
those in our universe (millions of years). For solar type stars, the
stellar lifetimes are shorter than those in our universe by about one
order of magnitude, so that a star with $M_\ast=1M_\odot$ burns its
hydrogen over $\sim1$ Gyr.  In addition, we find that luminosity
depends on stellar mass according to $L_\ast\propto M_\ast^2$, which
represents a less steep dependence than that for ordinary hydrogen
burning stars (where $L_\ast\propto M_\ast^3$). As a result, the
stellar lifetimes scale roughly as $t_\ast\propto1/M_\ast$. For the
upper branch of the main sequence, the longest-lived stars are those
with the minimum mass required to burn $^3$He into $^4$He, where this
mass scale is $M_\ast\approx0.03M_\odot$ (see Figures \ref{fig:lumvtime}
and \ref{fig:lumvtime18}). These stars live for trillions of years, much
longer than the expected lifetime of the Sun, and much longer than the
current age of our universe. The stars on the lower degenerate branch 
of the main sequence, with $M_\ast \simless10M_J$, can live even
longer, up to $\sim10^{14}$ yr. However, these stars are dim and cool,
with $L_\ast\sim10^{-7}-10^{-8}L_\odot$ and $T_\ast\sim300$ K.

\begin{figure}[tbp]
\centering 
\includegraphics[width=1.0\textwidth,trim=0 150 0 150]{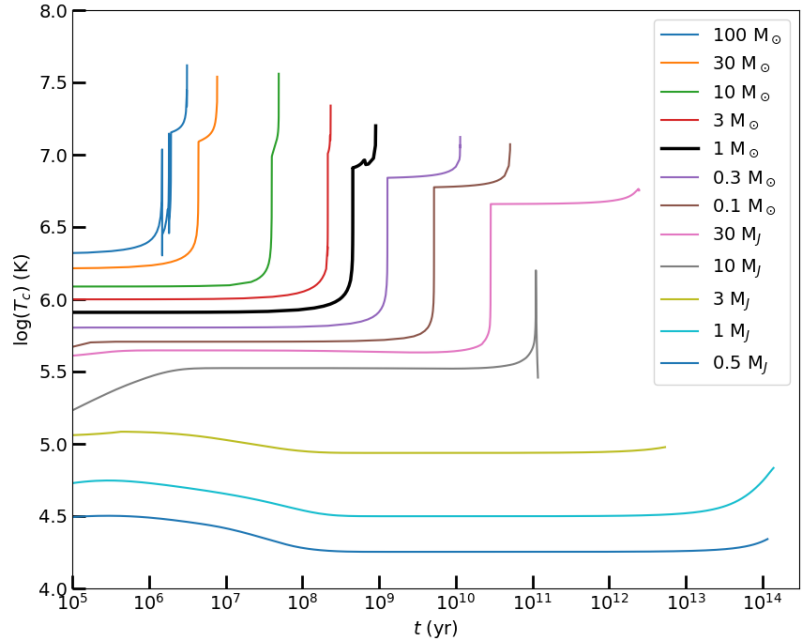}
\vskip 1.0truein 
\caption{Central temperature as a function of stellar age for stars 
with nuclear enhancement factor $X$ = $10^{15}$. } 
\label{fig:tempvtime} 
\end{figure}

\begin{figure}[tbp]
\centering 
\vskip 1.0truein 
\includegraphics[width=1.0\textwidth,trim=0 150 0 150]{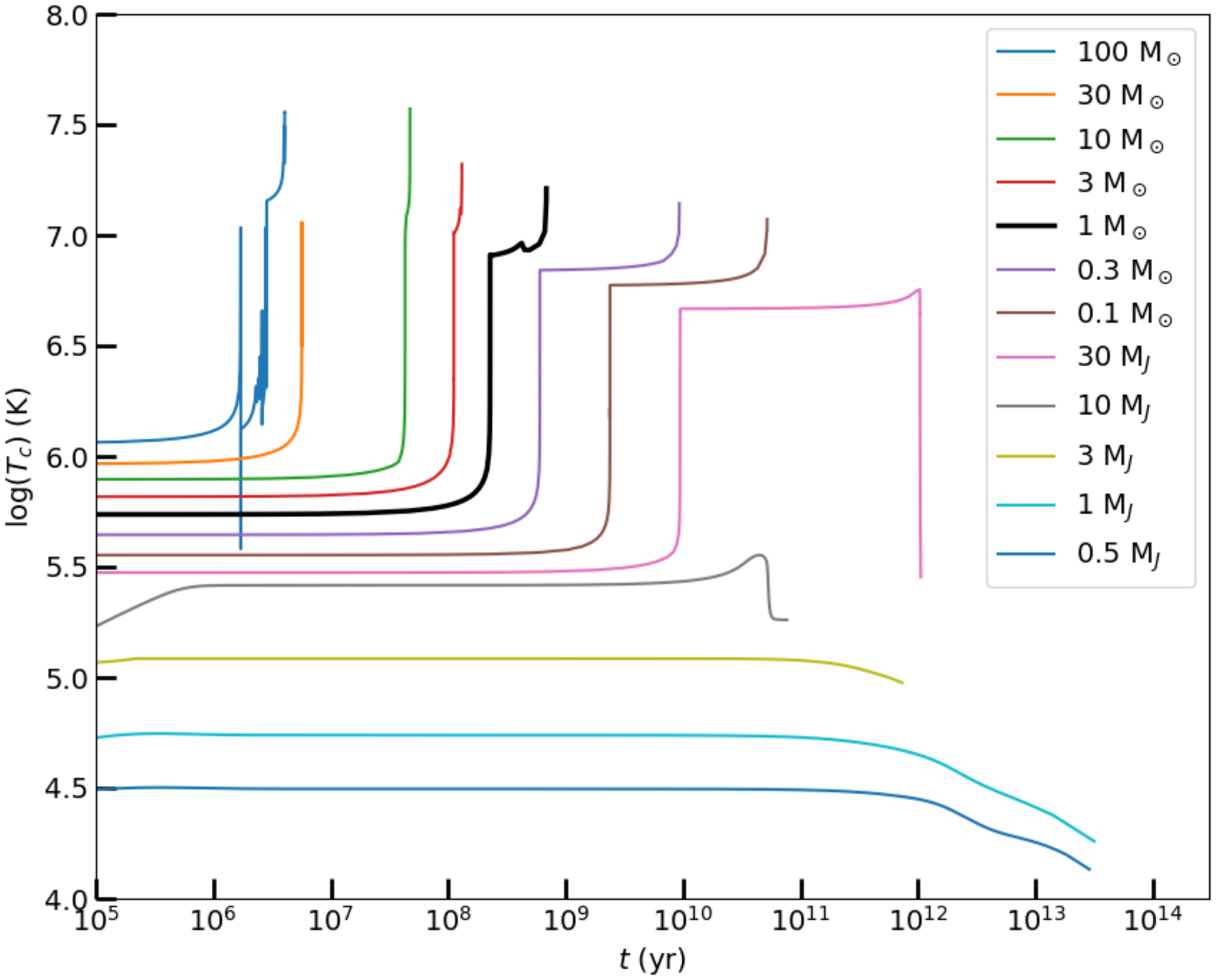}
\vskip 1.0truein 
\caption{Central temperature as a function of stellar age for stars 
with nuclear enhancement factor $X$ = $10^{18}$. } 
\label{fig:tempvtime18} 
\end{figure}

The central temperature is plotted as a function of stellar age in
Figures \ref{fig:tempvtime} and \ref{fig:tempvtime18} for the same
collection of stars (with nuclear enhancement factors $X=10^{15}$ and
$X=10^{18}$).  Results are shown for stellar masses ranging from
$M_\ast=0.5M_J$ up to $M_\ast=100M_\odot$. All of the stars display an 
extended phase of nearly constant central temperature corresponding to
burning of hydrogen into diprotons and then $^3$He. For higher mass
stars, this temperature is somewhat higher than $10^6$ K, whereas for
lower mass stars, the central temperature is somewhat smaller than
$10^6$ K (for stellar mass $M_\ast\simgreat10M_J$). For stars on the
lower portion of the main sequence, with $M_\ast<10M_J$, the central
temperature is substantially lower. All of these central temperatures
are below that required to process $^3$He into $^4$He (which
corresponds to $T_c\sim7\times10^6$ K), so the stellar cores are
converted into the lighter helium isotope.  The time span for this 
\propro-burning phase ranges from about 1 Myr for the largest stars ($M_\ast\sim100$ 
$M_\odot$), to 1 Gyr for solar type stars ($M_\ast\sim1M_\odot$), 
to 100 Gyr for the smallest stars on the upper main sequence 
($M_\ast\sim0.01M_\odot$). As discussed above, stars on the lower 
degenerate branch of the main sequence can live up to $10^{14}$ yr.

After the stellar cores exhaust their protons and attain a
largely $^3$He composition, the stars condense, the central
temperatures increases to $\sim10^7$ K, and $^3$He is processed into
$^4$He. For the higher mass stars, this phase has a comparable
lifetime to the earlier hydrogen burning phase (note the logarithmic
scale).  For lower mass stars, this second phase lasts even longer
than the initial phase, as long as the stars are massive enough to
burn $^3$He. The smallest stars, those with masses $M_\ast \simless$
0.025 $M_\odot$ have enough mass to produce $^3$He, but not to burn
it. As a result, the longest-lived stars (on the upper main sequence)
are those with the minimum mass required to burn $^3$He, with
$M_\ast\approx0.025M_\odot$, where the total lifetime is $\sim5.5$
trillion years. For comparison, the longest-lived stars in our
universe, with mass $M_\ast=0.08M_\odot$, have main-sequence lifetimes
of $\sim12$ trillion years \cite{lbastars}. Note that these timescales
are calculated for stars with solar metallicity,
$\metal=\metal_\odot$, and that lifetimes increase with increasing
metallicity (up to a maxmimum value $\metal\approx0.04>\metal_\odot$
\cite{alfuture}).

\begin{figure}[tbp]
\centering 
\includegraphics[width=1.0\textwidth,trim=0 150 0 150]{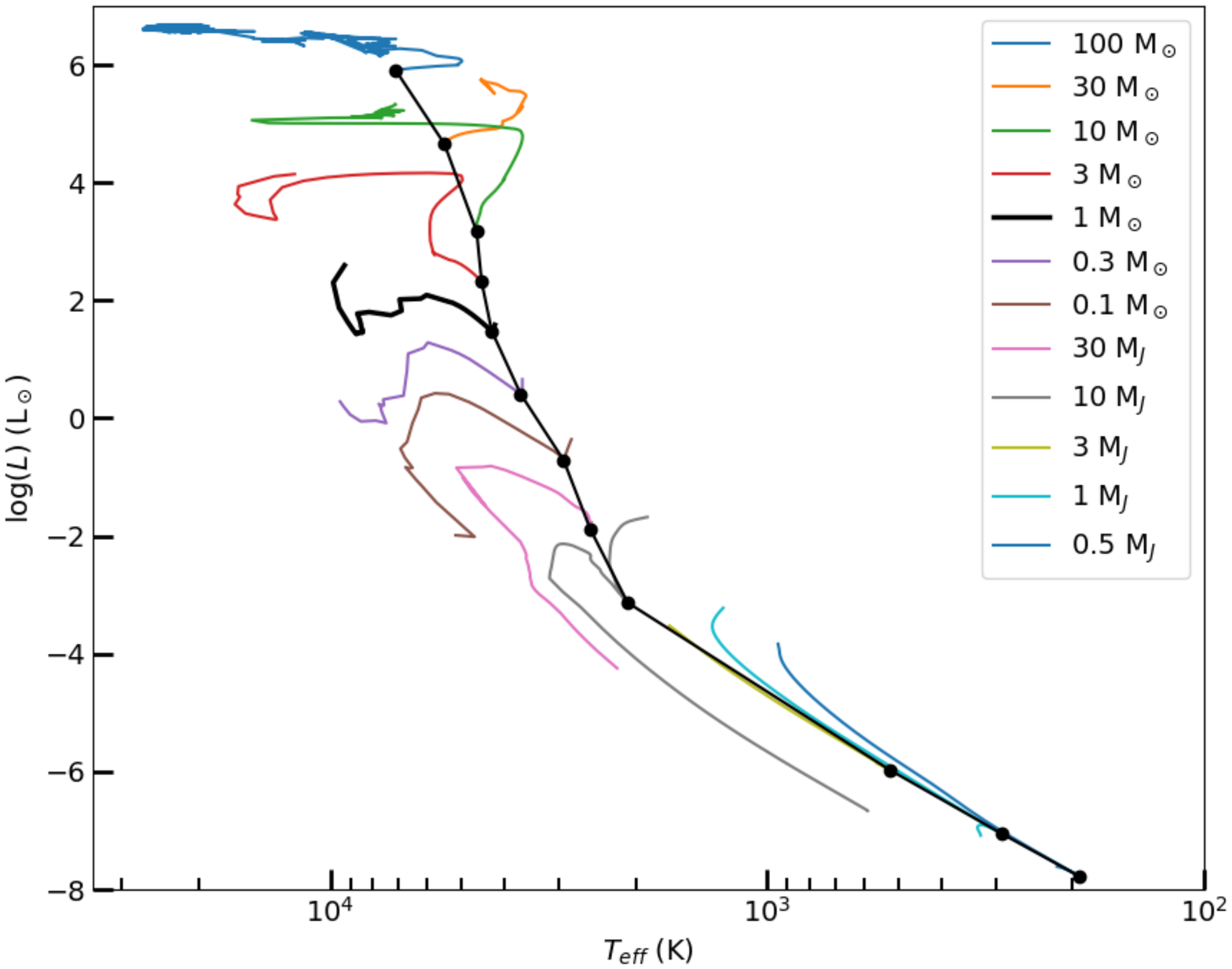}
\vskip 1.0truein 
\caption{Tracks in the H-R diagram for stars with stable diprotons
and nuclear enhancement factor $X=10^{15}$.  The black curve depicts
the Zero Age Main Sequence, ZAMS, defined here as the epoch when the
nuclear burning luminosity reaches 99\% of the total luminosity. The 
tracks to the right of the main sequence corresponds to the early 
pre-main-sequence evolution, whereas the tracks to the left correspond
to later post-main-sequence evolution. }
\label{fig:tracks15} 
\end{figure}

\begin{figure}[tbp]
\centering 
\includegraphics[width=1.0\textwidth,trim=0 150 0 150]{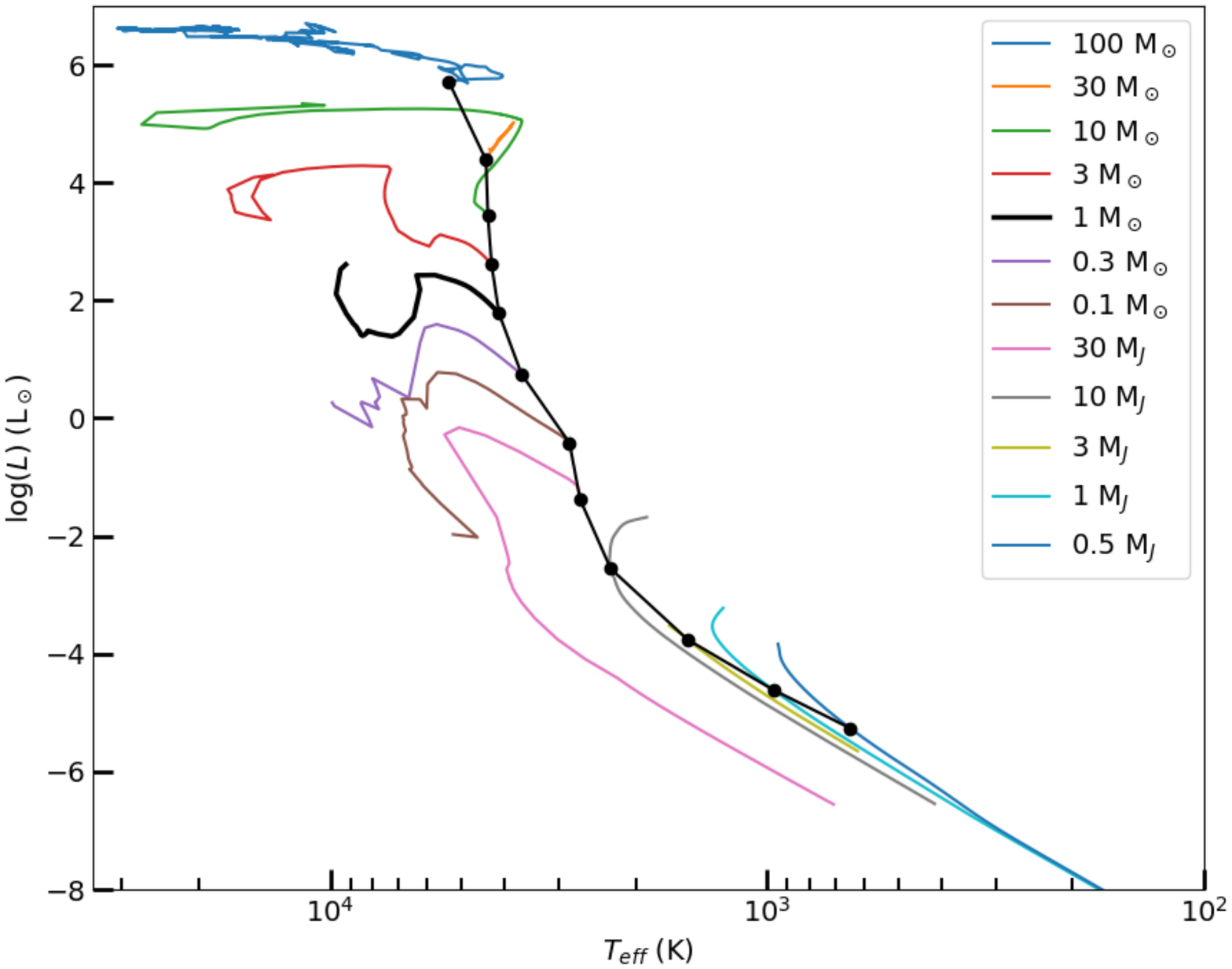}
\vskip 1.0truein 
\caption{Tracks in the H-R diagram for stars with stable diprotons
and nuclear enhancement factor $X=10^{18}$.  The black curve depicts
the Zero Age Main Sequence, ZAMS, defined here as the epoch when the
nuclear burning luminosity reaches 99\% of the total luminosity. The 
tracks to the right of the main sequence corresponds to the early 
pre-main-sequence evolution, whereas the tracks to the left correspond
to later post-main-sequence evolution. }
\label{fig:tracks18} 
\end{figure}

Figures \ref{fig:tracks15} and \ref{fig:tracks18} shows the tracks in
the H-R diagram for stars with bound diprotons over the same age range
used in the previous figures ($X=10^{15}$ and $X=10^{18}$,
respectively). The tracks start at stellar age $t=10^4$ yr, which is
somewhat shorter than their expected formation time, in order to see
how the stars evolve at early times. In this application, we use the
\mesa~ stellar evolution code to build stellar models with large radii
and then follow their subsequent evolution onto the main-sequence.
Future work should account for the star formation process, where the
bodies actively assemble their masses over time spans of $t_f\sim10^5$
yr \cite{adamsfatuzzo}. Although the initial decade in time should not
be considered as definitive, for stellar ages $t\simgreat10^5$ yr, the
stellar tracks reflect the physical behavior that is expected.  In
Figures \ref{fig:tracks15} and \ref{fig:tracks18}, the
pre-main-sequence (PMS) portions of the tracks in the H-R diagram are
confined to the right of the main sequence (lower surface
temperatures) and are relatively short (and would be shorter if they
were plotted after the isochrone for $10^5$ yr).

The PMS phase for diproton stars is thus shorter than that of ordinary
hydrogen burning stars, and this trend can be understood as follows: 
The central temperature is an increasing function of time as the stars
contract.  The central temperature ($T_c\sim10^6$ K) required for
diproton reactions is reached well before the temperature
($T_c\sim1.5\times10^7$ K) required for hydrogen burning, resulting in
the shorter PMS phase. When the central temperature reaches
$T_c\sim10^6$ K, and nuclear fusion begins, the vertical evolution in
the H-R diagram comes to a halt, and the stars lie along the zero-age
main-sequence (shown as the dark curves in Figures \ref{fig:tracks15}
and \ref{fig:tracks18}).

Figures \ref{fig:tracks15} and \ref{fig:tracks18} also show the post
main sequence evolution of the stars, up to the development of a
helium core. These tracks indicate that stars move to the left in the
H-R diagram (with increasing surface temperatures) as they exhaust
their hydrogen and $^3$He fuel. Subsequent stellar evolution, where
stars with $M_\ast\simgreat1M_\odot$ produce carbon and successively
larger nuclei, is expected to be the nearly same as in our universe
for a given stellar mass, and the corresponding tracks are not shown.
For the largest stars with $M_\ast=100M_\odot$, the tracks display
detailed structure, where the stars move back and forth in the diagram
during their post main sequence evolution (upper dark blue curves).
This complicated behavior arises because such large stars have enough
internal radiation pressure that they approach $n=3$ polytropes, which
are unstable \cite{clayton,kippenhahn,phillips}. Notice also that the
tracks for low-mass stars, with masses $M_\ast\simless0.3M_\odot$,
turn downward (with decreasing luminosity) at the end of time span
shown in the diagram. This behavior is expected: unlike solar-type
(and more massive) stars, low-mass stars do not become red giants
during their post main sequence evolution, and become bluer and dimmer
instead \cite{lbastars}.

\subsection{Mass-Luminosity Relation} 
\label{sec:masslum} 

In contrast to the standard mass-luminosity relation for radiative
hydrogen burning stars on the main sequence, where 
$L_\ast \propto M_\ast^3$, these numerical results show that diproton
stars have $L_\ast \propto M_\ast^2$. This trend holds over the entire
upper branch of the main sequence, for stellar masses in the range
$M_\ast$ = 0.01 -- 100 $M_\odot$.  The key features leading to this
form of the mass-luminosity relation are that both the central
temperature and the surface temperature are almost constant --- nearly
independent of stellar mass --- as discussed below.

First, consider the central temperature $T_c$. The enhanced cross
sections for nuclear interactions lead to lower operating temperatures
($T_c\sim10^6$ K). As a result, the classical turning point for
nuclear interactions is larger than for standard stellar interiors.
Because the particles interact through quantum mechanical tunneling,
which is exponentially suppressed, the reaction rate is extremely
sensitive to temperature. The lower temperature implies a larger
tunneling barrier, which makes the reaction rates an even more steeply
increasing function of $T_c$ (compared to main sequence stars in our
universe).  This extreme sensitivity, in turn, leads to a nearly
constant central temperature as a function of stellar mass.

The surface temperature $\teff$ is also nearly constant with varying
stellar mass, as shown by the main-sequence for diproton stars in the
H-R diagram (Figure \ref{fig:mainsequence}).  This behavior is related
to the Hayashi forbidden zone \cite{hayashi}, which prevents stars
from having surface temperatures that are too cool, and is caused
primarily by opacity effects. Although the derivation is somewhat
complicated, it can be shown that fully convective stars trace through
nearly vertical tracks in the H-R diagram \cite{kippenhahn,hansen}, so
that the surface temperature is nearly independent of luminosity and
slowly varying with stellar mass.  A simplified version of this
argument is presented in \ref{sec:contemp}.  These stars, which have
larger luminosities from the production of diprotons, are largely
convective and display this behavior.

Given central temperatures $T_c$ and photospheric temperatures
$\teff$ that are nearly constant, the mass-luminosity relation can
be understood as follows. The central pressure of the star must be
large enough to support the star against self-gravity and must be
provided by the ideal gas law. These two constraints can be written in
the approximate form 
\be 
P_c \approx {GM_\ast^2 \over R_\ast^4} \approx 
\rho_c {kT_c \over \mu} \approx {M_\ast \over R_\ast^3} 
{kT_c \over \mu} \,, 
\qquad {\rm or} \qquad kT_c \approx {GM_\ast \mu \over R_\ast} \,,
\ee 
where we ignore dimensionless factors of order unity and where $\mu$
is the mean particle mass in the stellar core. In the limit where
$T_c$ is independent of stellar mass, the above relations show that
$R_\ast \propto M_\ast$. Now consider the outer boundary condition,
\be 
L_\ast = 4\pi R_\ast^2 \sigma \teff^4 \,.  
\label{surfacebc} 
\ee 
For constant surface temperature, we find that 
\be 
L_\ast \propto R_\ast^2 \propto M_\ast^2 \,, 
\ee 
as found in the stellar evolution simulations. Putting in the physical
constants, we can derive an approximate expression for the stellar
luminosity as a function of mass: 
\be 
L_\ast = 4\pi \left({GM_\ast \mu \over kT_c}\right)^2 
\sigma \teff^4 \,.  
\label{lmsquare} 
\ee 
This relation leads to the simple power-law $L_\ast\sim M_\ast^2$ 
for constant $T_c$ and $\teff$. 

For completeness we note that the expression (\ref{lmsquare})
continues to be (approximately) valid even when the central
temperature $T_c$ and photospheric temperature $\teff$ vary with
mass. In actuality, both the $T_c$ and $\teff$ are slowly increasing
functions of stellar mass.  Moreover, the central temperature $T_c$
varies more rapidly than the surface temperature $\teff$. Figure
\ref{fig:tempvtime} shows that $T_c$ varies by a factor $\sim6.6$ over
the allowed mass range, so that $T_c^2$ varies by $\sim44$. For
comparison, Figure \ref{fig:mainsequence} shows that $\teff$ varies by
a factor of $\sim2.5$, so that $\teff^4$ varies by $\sim40$.  As a
result, the increase in $T_c^2$ in the denominator is effectively
offset by the increase in $\teff^4$ in the numerator, thereby leaving
the original approximation $L_\ast\sim M_\ast^2$ unchanged.

\begin{figure}[tbp]
\centering 
\includegraphics[width=1.0\textwidth,trim=0 150 0 150]{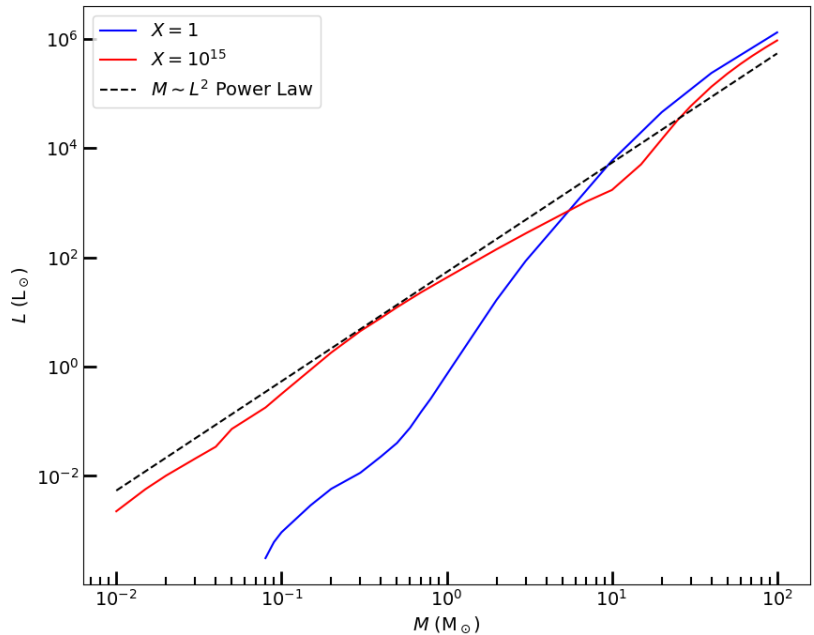}
\vskip 1.0truein 
\caption{Mass-luminosity relationship for stars in universes with 
stable diprotons (with enhancement factor $X=10^{15}$). The red curve
shows the mass-luminosity relation for the first stage of nuclear
burning, where protons are processed first into diprotons and then
into $^3$He. The dashed line shows the power-law relation from
equation (\ref{lmsquare}). For comparison, the blue curve shows the
mass-luminosity relation for main sequence stars in our universe. Both
sets of stellar models use solar metallicity. }
\label{fig:masslum} 
\end{figure}

Figure \ref{fig:masslum} shows the mass-luminosity relationship for
stars with stable diprotons on the upper branch of the main sequence
($M_\ast$ = 0.01 -- 100 $M_\odot$), where the numerical results are
compared to the analytic expression from equation (\ref{lmsquare}). 
The red curve shows the relation calculated using the \mesa~ code. 
The dashed line corrresponds to the $L_\ast\propto M_\ast^2$ relation
resulting from constant central temperature $T_c=10^6$ K and constant
photospheric temperature $\teff=3000$ K. The power-law approximation
works reasonably well over the four decades in stellar mass and
(nearly) nine decades in stellar luminosity shown in the figure. For
comparison, the blue curve shows the much steeper mass-luminosity
relationship applicable for stars in our universe (note the smaller
range in mass).

\subsection{Varying the Nuclear Enhancement Factor}
\label{sec:xvariation} 

In order to gain a clearer picture of the effects of the enhancement
factor $X$ on stellar evolution, this section considers a brief
exploration of the possible parameter space. Specifically, we simulate
stars with mass $M_\ast$ = 0.3 $M_\odot$ for a wide range of the
nuclear enhancement factors, $X$. Stars with this mass are sufficiently
long-lived to support complex life over a large range in $X$,
while still being massive enough to avoid the convergence issues
associated with lower masses in \mesa. Here we consider stars with
values of the enhancement factor $X$ from 10$^{-3}$ to 10$^{18}$. For
$X<10^{-3}$, stellar evolution is dominated by the CNO cycle even for
low-mass stars with $M_\ast$ = 0.3 $M_\odot$, so that further
reduction of the enhancement factor has negligible effect. For larger
enhancements, $X>10^{18}$, significant nuclear burning takes place
during the pre-main-sequence phase, so that the star formation process
(accretion history) must be included in the \mesa~
simulations.\footnote{As mentioned above, in this application, we do  
not simulate the formation of stars, but rather use \mesa~ to
construct pre-main-sequence models with arbitrarily large radii and
then let them evolve toward a hydrogen burning state. As a result, 
the earliest stages of evolution do not necessarily reflect the true
history of the star. Nonetheless, the models rapidly converge (in
time) towards physically realistic states. }

\begin{figure*}[htbp]
\includegraphics[width=0.99\textwidth]{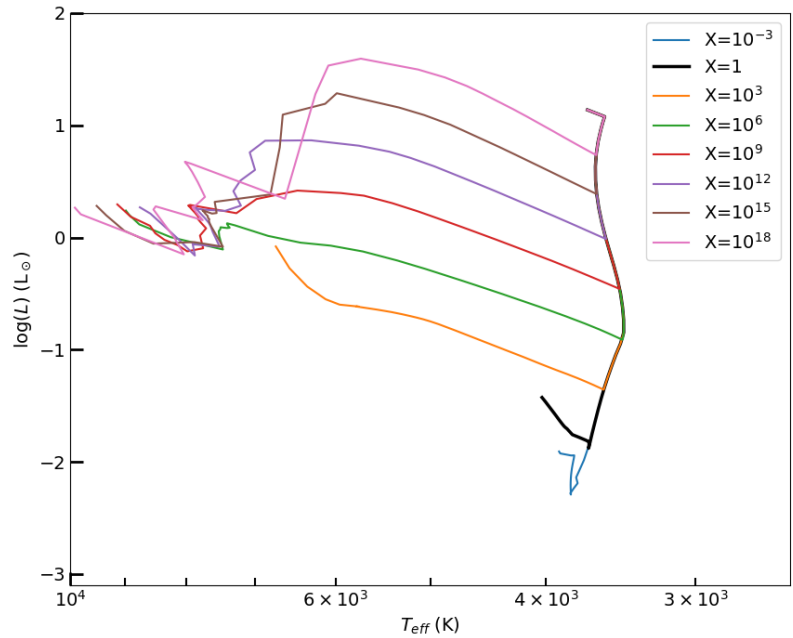}
\caption{Tracks in the H-R diagram for stars with $M_\ast$ = 0.3 
$M_\odot$ and varying values of the nuclear enhancement factor.  
The enhancement factor varies over the range $X=10^{-3}-10^{18}$, as
labeled. All of the stellar models show the same initial evolution as
they contract and move nearly vertically down the Hayashi track.
Contraction halts when the central temperature $T_c$ reaches that
required for nuclear burning, where the $T_c$ decreases (slowly) 
with increasing degrees of nuclear enhancement. }
\label{varC:HR}
\end{figure*}

\begin{figure*}[htbp]
\includegraphics[width=0.99\textwidth]{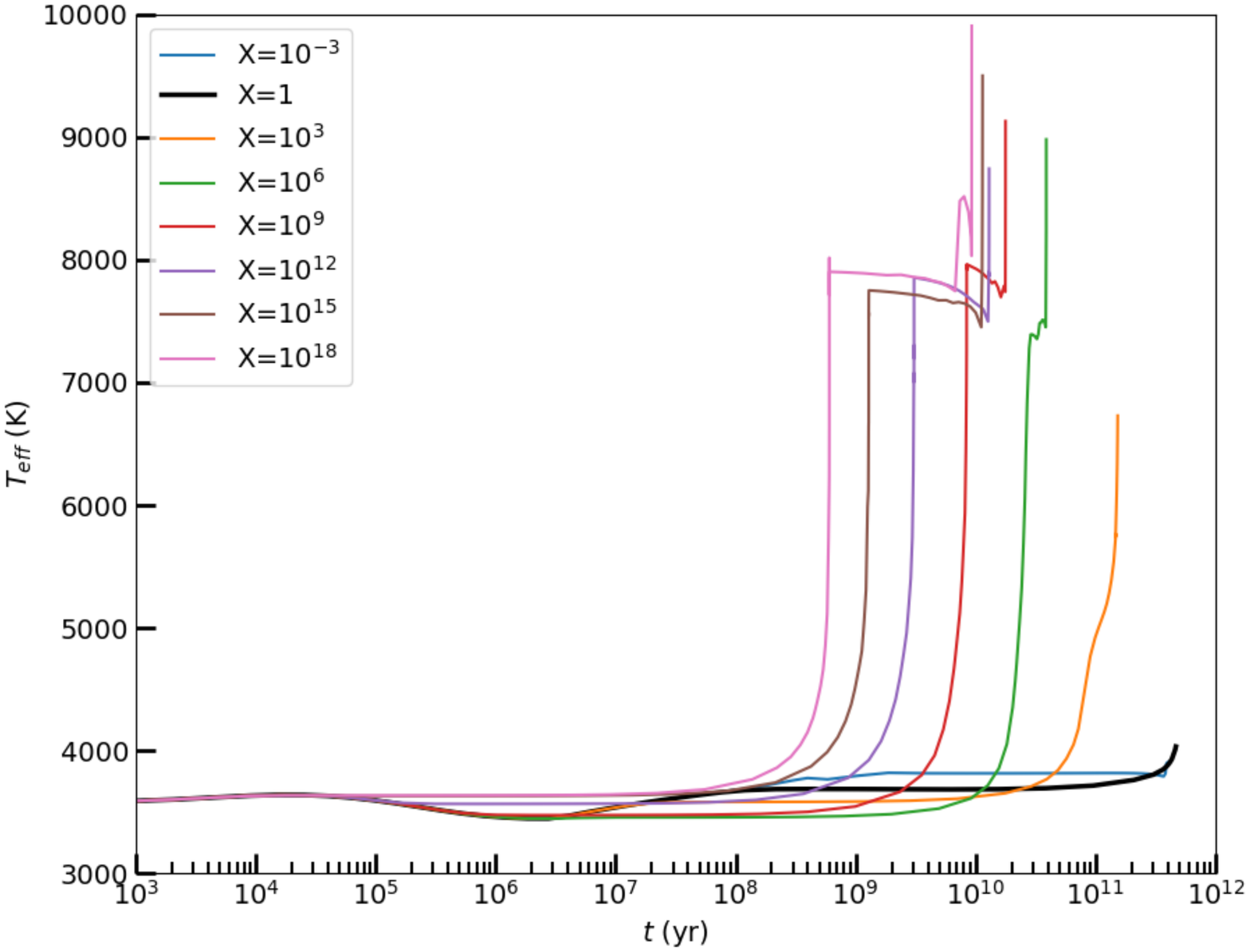}
\caption{Photospheric temperature $T_\ast$ as a function of time for 
stars with $M_\ast=0.3M_\odot$ and varying values of the nuclear
enhancement factor $X$ (as labeled). All of the stars have surface
temperatures $T_\ast\sim3500$ K during both their pre-main-sequence
phase and on their initial main sequence tracks, burning protons to 
diprotons and then $^3$He. For stars with $X\ge10^3$, the $^3$He
is not processed promptly into $^4$He, which is then produced later 
in separate phase with larger photospheric temperature. For stars in
our universe (and for $X<1$), $^4$He is produced promptly and the 
surface temperature remains low (black and blue curves). }  
\label{varC:Teff}
\end{figure*}

The evolution of $M_\ast=0.3M_\odot$ stars in the H-R diagram is
depicted in Figure \ref{varC:HR} over a range of 21 orders of
magnitude in the nuclear enhancement factor $X$. Since the stellar
mass is held constant, each star descends the same pre-main-sequence
(Hayashi) track until it reaches the central temperature necessary to
initiate fusion for a given level of enhancement. The star then
reaches the main sequence. For clarity, we also plot the effective
temperature $T_\ast$ as a function of time for these objects in Figure
\ref{varC:Teff}. The photospheric temperature remains nearly constant,
with $T_\ast\sim3500$ K, over the full range of enhancement factors,
although the lifetime decreases with increasing $X$. For stars with
$X\le1$, the hydrogen burning temperature is high enough that $^3$He
is burned into $^4$He at the same time. For larger values of $X\ge10^3$,
however, this latter reaction takes place later in a separate phase of
evolution, and the surface temperatures increase to
$T_\ast\sim7000-8000$ K.  Taken together, Figures \ref{varC:HR} and
\ref{varC:Teff} indicate that this blueward excursion in the H-R
diagram is associated with the transition from \propro fusion to
$^3$He fusion. Keep in mind that the initial \propro-burning stage
remains near the Hayashi track, as we saw with the $X=10^{15}-10^{18}$
cases above.

\begin{figure*}[htbp]
\includegraphics[width=0.99\textwidth]{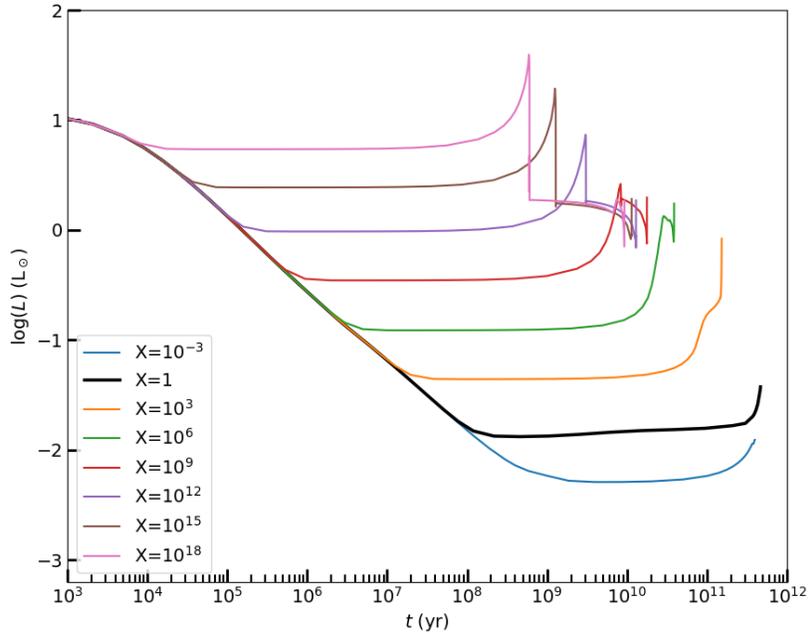}
\caption{Stellar luminosity $L_\ast$ as a function of time for stars
with $M_\ast=0.3M_\odot$ and varying values of the nuclear enhancement
factor (as labeled). The luminosity decreases with time as the stars
contract toward the main sequence, and then reaches a constant value 
when nuclear burning begins. The luminosity level increases slowly with
increasing nuclear enhancement factor, $X$. } 
\label{varC:Lum}
\end{figure*}

\begin{figure*}[htbp]
\includegraphics[width=0.99\textwidth]{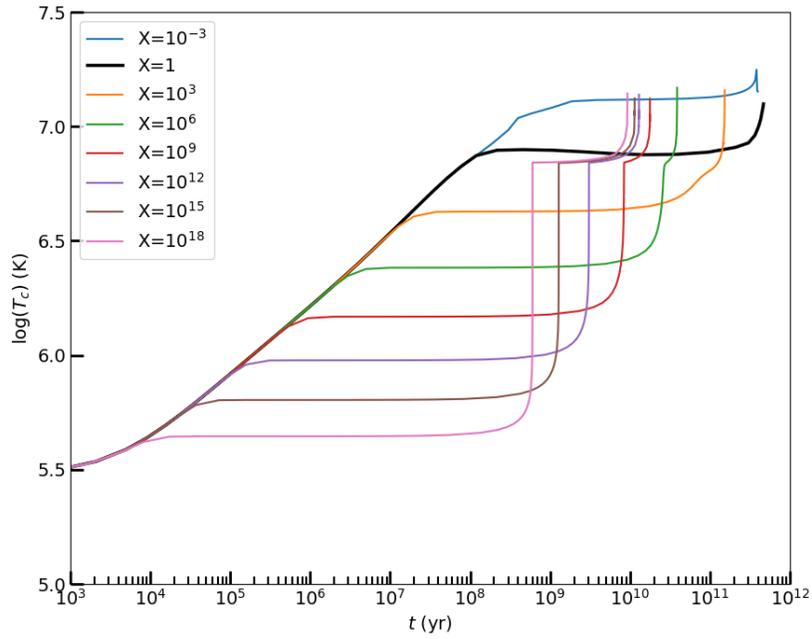}
\caption{Central temperature $T_c$ as a function of time for stars
with $M_\ast=0.3M_\odot$ and varying values of the nuclear enhancement
factor (as labeled). The central temperature increases with time until
nuclear burning begins, where the ignitition temperature decreases
with increasing values of the nuclear enhancement factor, $X$. For
stars in our universe (black curve), the nuclear burning temperature
is high enough that $^3$He is burned into $^4$He during the
(single) main sequence phase. For enhanced reaction rates, roughly
$X>10^3$, the central temperature of the first phase is below that
needed to burn $^3$He, which is processed during a later phase 
when $T_c$ abruptly increases to $\sim10^7$ K. }  
\label{varC:Tc}
\end{figure*}

To further elucidate the evolution of these objects, Figure
\ref{varC:Lum} shows the luminosity as a function of time, and Figure
\ref{varC:Tc} shows the central temperature versus time. These plots
cleanly demonstrate the (nearly) logarithmic dependence of luminosity,
central temperature, and stellar lifetime on the enhancement factor.
This dependence is specific to the \propro burning stage, which
corresponds to the extended phase of evolution with constant luminosity
and central temperature (for a given $X$). In contrast, the later
$^3$He-burning stage takes place through strong reactions, and is
therefore not significantly affected by the enhancement factor. This
logarithmic dependence further illustrates the robust nature of
stellar evolution in the face of changes in nuclear physics.

\subsection{Degenerate Very Low-Mass Stars }
\label{sec:vlm} 

One surprising result from our simulations is that nuclear burning
still occurs in very low-mass stars with masses $M_\ast\sim1-10M_J$,
below the deuterium-burning limit. These objects look significantly
different from other low-mass stars. With radii comparable to that of
Jupiter, these objects begin their evolution on cooling tracks similar
to brown dwarfs and eventually attain surface temperatures far below
the Hayashi limit for non-degenerate objects. The smallest of these
objects have surface temperatures $T_\ast< 300$ K, making them
``frozen stars'' where ice crystals could form in their atmospheres.
Moreover, the central temperatures of these objects are $T_c$ =
$3\times10^4-10^5$ K, which is normally considered too low to sustain
nuclear burning, even with strong (or electromagnetic) reactions.

It is possible that such unusual results are due to shortcomings in
the equation of state (or opacities) under the extreme conditions
realized in these objects. Nonetheless, a close inspection of the
simulations reveals that these objects are indeed physically
plausible. An order of magnitude treatment of the conditions required
for nuclear burning is carried out in \ref{sec:degenuke}. These
objects are partially degenerate, so that their hydrostatic
equilibrium is largely independent of nuclear reactions. In the
extreme limit, degeneracy can support the star against gravity, while
nuclear reactions take place in the background --- at an attenuated
rate --- provided that the nuclear enhancement factor (and hence the
reaction cross section) is large enough. The enormous densities
associated with degenerate objects lead to strong electron screening,
which negates some of the Coulomb repulsion and allows for nuclear
reactions to occur at lower central temperatures.  Nuclear burning is
often described as ``switching on'' at some specified
temperature. Yet, although the reaction rates are exponentially
dependent on temperature, they are always taking place at some low
level.  Non-degenerate objects cannot be supported in such a regime,
and must adjust to another stellar configuration. However, the very
low-mass objects considered here are largely supported without their
nuclear contribution, so that these objects can generate energy at
anemic rates. With their low central temperature, the associated
nuclear burning rate is extremely slow, such that the stellar lifetime
is longer than $10^{14}$ years, despite the lower yield of burning to
$^3$He rather than $^4$He. With low luminosity and fixed radius, these
objects can thus maintain surface temperatures $T_\ast\sim$300 K.

The strange properties of these hypothetical very low-mass stars stem
from them being partially degenerate, like brown dwarfs.  Under these
conditions, the central region of the star (where
nuclear burning occurs) does not need to reach a pressure equilibrium
to support the star against gravity, but is still able to reach and
maintain a {\it thermal} equilibrium.  This latter condition is
reached at a much lower temperature (compared with that required for
pressure equilibrium). After $\sim10^9$ yr, the nuclear
burning luminosity dominates over heating from Kelvin-Helmholtz
contraction, and the ``star'' can function as a nuclear-burning entity.

Note that the possibility of this unusual stellar state is a result 
of strong-burning nuclear reactions in general rather than the 
$p(p,\diproton)\gamma$ reaction in particular. Intriguingly, this
finding suggests that if Jupiter were made of pure deuterium in our
own universe, it could also reach a thermal equilibrium of nuclear
burning. It also suggests that similar very low-mass stars could occur
in a ``weakless'' universe \cite{weakless}. In previous work
\citep{grohsweakless}, we found that in some universes without a weak
interaction, stars will be composed mostly of deuterium and will
produce energy via strong reactions, but we did not explore stars
smaller than the standard deuterium burning limit. Nonetheless, very 
low-mass degenerate stars in such a universe could operate by the same
mechanism.

The minimum mass of these degenerate stars depends on the enhancement
factor, but appears to be significantly less than $\sim1M_J$. We were
not able to definitively determine this minimum mass from our
simulations because of convergence problems at low masses (primarily
due to the equation of state for the extreme conditions associated
with these very low-mass stars). The smallest star that we were able
to successfully simulate had mass $M_\ast$ = 0.5 $M_J$. However,
simulations run with a range of enhancement factors suggest that the
minimum stellar mass lies in the neighborhood of $M_\ast$ = 0.25 $M_J$
for $X=10^{15}$.  For lower masses, the nuclear burning luminosity
never exceeds the Kelvin-Helmholtz luminosity, and the object evolves
as an ordinary sub-brown dwarf or giant planet. 


\section{Conclusion} 
\label{sec:conclude} 

This paper has considered stellar evolution in universes with stable
diprotons. In this scenario, nuclear reactions can take place in
stellar cores through the strong force, with reaction rates enhanced
by factors $X\sim10^{15}-10^{18}$. Our main results are summarized in 
Section \ref{sec:summary}, with a discussion of their implications 
presented in Section \ref{sec:discuss}. 

\subsection{Summary of Results} 
\label{sec:summary} 

The most important result of this study is that stars continue to
operate normally with large enhancements of their nuclear reaction
cross sections, such as those expected in universes with stable diprotons. The
stellar luminosities, lifetimes, and surface temperatures are roughly
similar to those of stars in our universe.  Although some differences
arise, as described below, these stars do not have catastrophically
short lifetimes (as many previous papers have asserted, e.g.,
\cite{dyson,davies1972,bartip,pochet,tegmark1998,hogan,reessix,
dentfair,tegmark2006}). The longest-lived stars can sustain nuclear 
fusion for trillions of years, about three orders of magnitude longer
than the current age of our universe.

Considerations of Big Bang Nucleosynthesis with bound diprotons
(Section \ref{sec:bbn}) indicate that the abundances of protons,
deuterium, and helium will be approximately the same as in standard
BBN. The production of diprotons is suppressed relative to deuterium
because of the coulomb barrier, in spite of the enhanced cross
section. The diprotons that are produced will eventually be transmuted
into deuterium, thereby raising its abundance. Nonetheless, we expect
$Y_d \ll Y_p$, so that the starting conditions for stellar evolution 
and nucleosynthesis are largely unchanged (see also 
\cite{bradford,macdonaldmullan}). 

The main sequence for stars with stable diprotons can be separated
into two branches (Figure \ref{fig:mainsequence}). The upper main
sequence corresponds to stellar masses $M_\ast=0.01-100M_\odot$.
Compared to stars with standard nuclear reaction rates, these stars
are somewhat more luminous and have redder
surfaces. As a result, the main sequence for these objects is steeper
than the standard one, with surface temperatures spanning the narrower
range $\teff\sim2500-5500$ (see Figure \ref{fig:mainsequence}). The
largest stars (with $M_\ast=100M_\odot$) have approximately the same
luminosity as those in our universe, whereas lower mass stars have
higher luminosity. As a result, the luminosity range is compressed for
the same span of stellar masses.  However, this trend is compensated for
by the larger range in stellar mass that can sustain nuclear
burning. Over the upper main sequence, with masses
$M_\ast=0.01-100M_\odot$, the overall range in stellar luminosity is
roughly comparable to that of our universe, i.e.,
$L_\ast\sim10^{-3}-10^6L_\odot$. The solar-type stars, with somewhat
cooler surfaces and more luminosity than our Sun, have configurations
roughly similar to those of red giants in our universe.  Note that
this same set of stellar properties arises for stars that are
primarily composed of deuterium \cite{grohsweakless,howeweakful},
which has nuclear reaction rates comparable to those of diprotons.

In addition to stars that are supported by the pressure resulting from
their internal nuclear reactions, the greatly enhanced reaction rates
allow for a new type of hybrid stellar configuration that corresponds
to the lower ``degenerate branch'' of the main sequence shown in Figure \ref{fig:mainsequence}.
Stellar objects with masses $M_\ast=1-10M_J$ can be supported (in
part) by degeneracy pressure, but still sustain nuclear reaction rates
high enough to produce quasi-stable surface temperatures $T_\ast\sim300$ K. These
objects are like brown dwarfs in our universe in that they do not
generate enough power to be supported through thermal pressure. On the
other hand, the residual nuclear reactions are effective enough that
that the bodies can maintain a nearly constant luminosity over
extended time scales (see Section \ref{sec:vlm} and
\ref{sec:degenuke}).

For masses comparable to the Sun, the lifetimes for stars with stable
diprotons are shorter than those burning hydrogen by about one order
of magnitude (see Figures \ref{fig:lumvtime} --
\ref{fig:tempvtime18}). low-mass stars live much longer than larger
stars, so that stars with $M_\ast\approx0.3M_\odot$ have main-sequence
lifetimes comparable to solar-type stars with standard nuclear
reactions. Moreover, because of the much larger cross sections for
diproton reactions, stars in such universes can sustain nuclear
reactions with much smaller masses than in our universe, so that the
upper main sequence extends down to masses $M_\ast\approx0.01M_\odot$,
with commensurately longer lifetimes. 
These results vary logarithmically with the enhancement factor $X$
(see Section \ref{sec:xvariation} and \ref{sec:timevx}).

The range of stellar masses for the upper main sequence thus spans a
factor of $\sim10^4$, with an additional decade in mass contributed by the lower
main sequence, compared with a mass range of $\sim10^3$ in our universe.
With a larger range in mass, and a similar range in luminosity, the
mass-luminosity relation over the upper main sequence is less steep
than that of our universe. To a good approximation, we find
$L_\ast\propto M_\ast^2$ (see Figure \ref{fig:masslum}), and this
result can be understood in terms of basic stellar physics (see Section
\ref{sec:masslum} and \ref{sec:contemp}).

With greatly enhanced reaction cross sections, the usual \propro chain
of nuclear reactions takes place in two stages.  In the first stage,
protons fuse into diprotons, which capture electrons to become
deuterium and then interact with protons to become $^3$He. Because of
the lower nuclear burning temperature for this process, the fusion of
$^3$He into alpha particles does not take place promptly, but rather
occurs later during the second stage. The smallest stars can sustain
nuclear fusion and produce $^3$He, but do not have enough mass to
process the $^3$He into $^4$He (which requires a higher central
temperature). As a result, the longest-lived (non-degenerate) 
stars in universes with stable diprotons have masses
$M_\ast\approx0.025M_\odot$, the minimum needed to produce $^4$He,
and live for $t\sim6000$ Gyr.  This
lifetime is comparable to that of the longest-lived stars in
our universe ($M_\ast=0.08M_\odot$ and $t\sim10^{4}$ Gyr) and much
longer than the current cosmic age ($\sim14$ Gyr).  The partially
degenerate stars on the lower main sequence can sustain nuclear
processes even longer, up to $t\sim10^{5}$ Gyr.

\subsection{Discussion}  
\label{sec:discuss} 

The main result of this paper is that stellar evolution is only
modestly affected by the greatly enhanced nuclear reaction cross
sections expected in universes with stable diprotons. In particular,
the smallest such stars can still live for trillions of years, far beyond
the current age of the universe. Since this finding is in stark
contrast to many previous claims in the literature, it is useful to
eludicate why stars are so impervious to changes in the input nuclear 
physics.

First, note that stellar lifetimes do not vary inversely with the
nuclear reaction cross sections, but only logarithmically. The energy generation rate in the
stellar core must compensate for the energy lost from the stellar
surface, but stars can adjust their reaction rates by expanding or
contracting. Since pressure is provided by the ideal gas law, changes
in stellar radius lead to corresponding changes in the core
temperature. The nuclear reaction rates are exponentially sensitive to
temperature, so that a small decrease in $T_c$ can compensate for an
enormous increase in the cross section. Specifically, for enhancement
factors in the range $X=10^{15}-10^{18}$, the operating temperature of
stars decreases from $T_c\approx1.5\times10^7$ K (the value for our
Sun) down to about $T_c\sim10^6$ K. This modest change in $T_c$ lowers
the nuclear reaction rate by a factor comparable to the enhancement
factor $X$ of the cross section. The reason for the extreme
sensitivity is that the nuclear reactions take place via quantum
mechanical tunneling, and the barriers are substantial. In the core of
the Sun, for example, the typical proton separation (24,000 fm) and
the classical turning point ($\sim1000$ fm) are both much larger than
the range of the strong force ($\sim10$ fm). The protons thus have a 
formidable barrier to tunnel through. 

The characteristic temperature of $T_c\sim10^6$ K for stars with stable
diprotons is essentially equivalent to the well-known deuterium
burning temperature \cite{dburntemp}. This coincidence is not
surprising, because the reaction cross section for deuterium burning
is larger than that for the standard \propro chain by a factor of
$\sim10^{18}$, and deuterium reactions are the strong-force analog
reactions used to estimate the enhancement factors for diproton
reactions.

In our universe, most stars experience a deuterium burning
phase during their early evolution. Stars are born with radii too
large and central temperatures too low to sustain hydrogen
fusion.\footnote{This discussion applies only to stars with masses
$M_\ast\simless$ 7 $M_\odot$. The largest stars, corresponding to
the most massive $\sim1\%$ of the stellar population, burn their
deuterium as they form.} As a result, newly born stars are not
powered by nuclear reactions, but rather by gravitational
contraction. During this pre-main-sequence phase, the stellar radius
decreases and the central temperature grows. When the temperature
becomes becomes high enough, $T_c\sim10^6$ K, deuterium burning
commences and delays further contraction. Because the deuterium
abundance is low, D/H $\sim2\times10^{-5}$, this phase only lasts for
$\sim10^5$ years before the deuterium fuel is exhausted in the stellar
core. Some time later, 0.1 -- 10 Myr depending on the stellar mass,
the core reaches the higher temperature ($T_c\sim1.5\times10^7$ K)
necessary for hydrogen burning. In universes with bound diprotons and
enhanced nuclear cross sections, however, stars can burn through the
bulk of their fuel with central temperatures $T_c\sim10^6$ K. Since
the nuclear fuel is more abundant by a factor $\sim10^5$ (compared to
the deuterium supply in our universe), solar type stars can sustain
nuclear fusion for billions of years.

In this consideration of universes with bound diprotons, we are
assuming that the strong force is somewhat stonger than in our
universe (with an increase of order $\sim10\%$, 
\cite{davies1972,bartip,hogan,reessix}). However, apart from the
existence of the bound state $\diproton$ and enormous enhancements of
the cross sections ($X=10^{15}-10^{18}$), we are implicitly assuming
that the rest of nuclear physics is largely unchanged. In particular,
the particle spins are the same as in our universe, and the ordering of
nuclear binding energies remains constant. These properties imply that
nucleons will cascade into larger nuclei as in our universe, with the
progression 
\be 
p \quad \longrightarrow \quad (\,\diproton\,\,\to\,\,d\,) 
\quad \longrightarrow \quad ^3{\rm He} \quad \longrightarrow 
\quad ^4{\rm He} \,.  
\ee 
One expects the weak force to vary along with the strong force, but
the modest changes considered here will not affect stellar operations.
If anything, the key process of electron capture from equation
(\ref{dipcapture}) will be even more effective. 

Although the expected small changes in binding energy do not affect
the progression from protons to helium outlined above, the later
stages of nuclear burning in massive stars could be altered. In this
regime, relatively small changes in binding energies could lead to
interesting systematic differences \cite{clayton,kippenhahn}. As
nuclei of ever larger atomic number undergo nuclear burning, the time
scales decrease (down to minutes and even seconds), and the reaction
networks become increasingly complicated. As a result, the relative
abundances of heavy elements (up to iron and beyond) produced by
massive stars could be somewhat different in universes with stable
diprotons. This issue is beyond the scope of this present paper and is
left for future work.

One should keep in mind that much larger changes to the strong force
can affect nuclear structure and the manner in which stars evolve.
This issue is complicated by the nature of the strong interaction,
which is ultimately determined through QCD, but can be described by an
effective nuclear potential with an overall strength, a length scale
(range), and a repulsive core (e.g., see \cite{chiral} and references
therein).  Quantum statistics also plays an important role, especially
for light nuclei, as bound states must be overall antisymmetric in
spin and isospin.  All of these components to the potential will
affect nuclear structure.  As an extreme example, if the strong force
becomes much stronger, then nucleons become relativistic in their
bound nuclear states. In addition, the nuclei could transform to
states of quark matter, where free quarks provide the degrees of
freedom instead of protons and neutrons. The required variations to
the strong force potential necessary to instigate these changes has
not been determined. Since nuclear binding energies are of order 1 MeV
per particle, and the QCD phase transition temperature is $\sim200$
MeV (and nucleon masses are $\sim1$ GeV), an increase in the strong
force by a factor of $\sim100$ could lead to vastly different nuclear
structures. Even in this case, the nuclei could remain as bound
entities, and their role in atomic physics would be essentially
unchanged. However, the astrophysical processes that produce the
nuclei, in BBN and stellar nulceosynthesis, could be quite different.
(For example, as noted in Section \ref{sec:xvariation}, if these
changes result in enhancement factors $X\simgreat10^{18}$, nuclear
burning would be able to occur during star formation.)  These issues
should be explored in future work.

This study demonstrates that stars are much less sensitive to
variations in the fundamental parameters of physics and astrophysics
than is often claimed. This work shows that stars operate normally
even with enormous enhancements (15 -- 18 orders of magnitude) in
their nuclear reaction cross sections, as expected in universes with
stable diprotons (see also \cite{adamsreview,barnes2015}). Previous
work has shown that stars can maintain stable nuclear burning
configurations with the fine structure constant a factor of $\sim100$
larger or smaller, and with an even larger possible range for the
gravitational constant \cite{adams2008,adams2016}. Stars can also
function in universes with no weak interactions \cite{grohsweakless},
stronger weak interactions \cite{howeweakful}, stable beryllium-8
\cite{agalpha}, over a large range of carbon-12 resonances
\cite{huang,livio,schlattl,epelbaum}, and even in universes without
stable deuterium \cite{agdeuterium} (cf. \cite{barnes2017}).  Taken
together, these findings indicate that stars are not the limiting
factor for universes to remain habitable.

\vskip0.15truein

\noindent
{\bf Acknowledgments:} We are grateful to Martin Rees and Frank Timmes
for useful feedback.  The work of FCA is supported in part by NASA
Grant NNX16AB47G and by the Leinweber Center for Theoretical Physics
at the University of Michigan. AH is supported by an appointment to
the NASA Postdoctoral Program at the NASA Goddard Space Flight Center,
administered by Universities Space Research Association under contract
with NASA. EG and GF acknowledge additional support from the National
Science Foundation, Grant PHY-1630782, and the Heising-Simons
Foundation, Grant 2017-228. GF also acknowledges NSF grant PHY-1914242
at University of California San Diego.

\vskip0.10truein

\appendix
\section{Additional Nuclear Reactions} 
\label{sec:newnukes}  

The existence of bound states of two protons ($\diproton$) and/or two
neutrons ($\dineutron$) requires the inclusion of new nuclear
reactions. The possible reactions for light elements are listed in
Table \ref{table:newnukes} and are organized into different
categories. The cross sections --- and hence the reaction rates ---
for these new processes are expected to be comparable to those of
analog reactions that exist in our universe (also shown in the
table). For the reactions that include diprotons, we can add a neutron
to one of the reactants on both sides of the equation and obtain an
analog reaction with a known/measured reaction rate. The diproton
becomes $^3$He under this procedure.  Similarly, for reactions that
involve the dineutron, we can add a proton to both sides of the
reaction to obtain an analog process. In this case, the dineutron
becomes tritium. We assume here that the analog reactions, which have
known cross sections, will provide order of magnitude estimates for
the new reactions. Note that the cross sections depend on the spin,
charge, and mass of the participating particles. As a result, the
analog cross sections can be different from the values for the new
reactions by factors of order unity. For completeness we note that for
the analog reactions involving tritium ($t$), the reacting particles
have increased charges and hence different Gamow factors. At low
energies, the analog tritium reactions will thus be suppressed
relative to the diproton/dineutron reactions (although these reactions
do not play an important role in the stellar evolution calculations of
this paper).

\begin{table} 
\centerline{\bf Nuclear Reactions for Diprotons and Dineutrons $\qquad$} 
\vskip8pt
\begin{tabular}{lll} 
\hline
\hline
Reaction Type & New Reaction & Analog Reaction \\ 
\hline \hline 
Production & $p~(p,\gamma)~\diproton$ & $d~(p,\gamma)~^3{\rm He}$ \\
 & $n~(n,\gamma)~\dineutron$ & $d~(n,\gamma)~t$ \\ 
\hline
Weak & $\diproton~(e^{-},\nu_e)~d$ & $^3{\rm He}~(e^{-},\nu_e)~t$\\ 
 & $\dineutron~(e^{+},{\bar\nu}_e)~d$ & $~t~(e^{+},{\bar\nu}_e)~^3{\rm He}$\\
\hline
Single Nucleon & $\diproton~(n,p)~d$ & $^3{\rm He}~(n,d)~d$ \\ 
 & $\diproton~(n,\gamma)~^3{\rm He}$ & $^3{\rm He}~(n,\gamma)~^4{\rm He}$ \\
 & $\dineutron~(p,n)~d$ & $~t~(p,n)~^3{\rm He}$ \\ 
 & $\dineutron~(p,\gamma)~t$ & $~t~(p,\gamma)~^4{\rm He}$ \\ 
\hline
Diproton + Dineutron &  
$\diproton~(\dineutron,\gamma)~^4{\rm He}$ & 
$^3{\rm He}~(t,\gamma)~^6{\rm Li}$ \\ 
 & $\diproton~(\dineutron,n)~^3{\rm He}$ & $^3{\rm He}~(t,d)~^4{\rm He}$ \\ 
 & $\diproton~(\dineutron,p)~t$ & $^3{\rm He}~(t,d)~^4{\rm He}$ \\ 
 & $\diproton~(\dineutron,d)~d$ & $^3{\rm He}~(t,t)~^3{\rm He}$ \\ 
\hline
Deuteron & $\diproton~(d,p)~^3{\rm He}$ & 
$^3{\rm He}~(d,p)~^4{\rm He}$ \\ 
 & $\dineutron~(d,n)~t$ & $~t~(d,n)~^4{\rm He}$ \\ 
\hline
$A=3$ Targets & $~t~(\diproton,p)~^4{\rm He}$ & 
$^3{\rm He}~(t,d)~^4{\rm He}$ \\ 
 & $~t~(\diproton,d)~^3{\rm He}$ & $^3{\rm He}~(t,t)~^3{\rm He}$ \\ 
 & $^3{\rm He}~(\dineutron,n)~^4{\rm He}$ & $^3{\rm He}~(t,d)~^4{\rm He}$ \\ 
 & $^3{\rm He}~(\dineutron,d)~t$ & $^3{\rm He}~(t,t)~^3{\rm He}$ \\ 
\hline 
Helium Targets & $^4{\rm He}~(\diproton,\gamma)~^6{\rm Be}$ & 
$^4{\rm He}~(^3{\rm He},\gamma)~^7{\rm Be}$ \\ 
 & $^4{\rm He}~(\diproton,~^3{\rm He})~^3{\rm He}$ & 
$^4{\rm He}~(^3{\rm He},~^3{\rm He})~^4{\rm He}$ \\ 
 & $^4{\rm He}~(\dineutron,\gamma)~^6{\rm He}$ & 
$^4{\rm He}~(t,\gamma)~^7{\rm Li}$ \\ 
 & $^4{\rm He}~(\dineutron,t)~t$ & 
$^4{\rm He}~(t,t)~^4{\rm He}$ \\ 
\hline \hline  
\end{tabular} 
\caption{Nuclear reactions involving diprotons and dineutrons. The
reactions can be grouped according to the type of particles involved
as specified in the first column. New reactions involving diprotons
and dineutrons are listed in the second column. Analog nuclear
reactions that operate in our universe, and which are expected to 
have comparable cross sections, are listed in third column. } 
\label{table:newnukes} 
\end{table}  
 
\section{Photospheric Temperatures for Convective Stars}  
\label{sec:contemp} 

This Appendix shows that fully convective stars have nearly constant
surface temperature as a function of stellar mass.  For the stars of
interest, the enhanced interaction cross sections lead to somewhat
larger luminosities, and the convective approximation provides a good
model. These stars can be described by $n=3/2$ polytropes, where the 
equation of state takes the form  
\be
P = K \rho^{5/3} \,.
\ee
This equation of state can be rewritten in terms of temperature 
\be
P = C (kT)^{5/2} \,,
\label{condef} 
\ee
where we have assumed that the ideal gas law holds. Since the star is
a complete polytrope, the constant $C$ has the same value throughout
the stellar interior, and can be written in terms of the mass and
radius \cite{hansen}, i.e., 
\be
C = \alpha \, G^{-3/2} \mu^{-5/2} M_\ast^{-1/2} R_\ast^{-3/2} \,, 
\label{conspec} 
\ee
where $\mu$ is the mean molecular weight of the gas. The 
dimensionless constant $\alpha$ is independent of the mass 
and radius $(M_\ast,R_\ast)$.

The pressure at the photosphere can be written in the form 
\be
P_\ast = {2 \over 3} {g_\ast \over \kappa_\ast} = 
{2 \over 3 \kappa_\ast} {GM_\ast \over R_\ast^2} \,,
\label{psurface} 
\ee
where $\kappa_\ast$ is the photospheric opacity. This expression
follows from assuming hydrostatic equilibrium and using the fact that
the optical depth $\tau$ = 2/3 at the photosphere. We can combine the
expressions (\ref{condef}) and (\ref{psurface}) for the photospheric
pressure, along with the definition (\ref{conspec}), to find 
\be
P_\ast = {2 \over 3 \kappa_\ast} {GM_\ast \over R_\ast^2} 
= \alpha \, G^{-3/2} \mu^{-5/2} M_\ast^{-1/2} R_\ast^{-3/2} 
(k\teff)^{5/2} \,. 
\ee
After some rearrangement this expression becomes
\be
(k\teff)^{5/2} \kappa_\ast(\teff) = 
{2 \over 3\alpha} G^{5/2} \mu^{5/2} 
M_\ast^{3/2} R_\ast^{-1/2} \,,
\ee
which shows how the temperature dependent quantities scale with
stellar mass, radius, and composition. Now consider the standard case
where the opacity at the photosphere is extremely sensitive to the
temperature. The opacity is generally dominated by interaction with
H$^{-}$, and this opacity has an approximate temperature dependence of
$\kappa_\ast \sim T^9$. If we use this power-law form for the opacity,
we obtain the scaling law
\be 
\teff \propto \mu^{5/23} M_\ast^{3/23} R_\ast^{-1/23} \,.
\label{tvmr} 
\ee
The weak dependence on the stellar parameters $(M_\ast,R_\ast)$ 
thus shows that the photospheric temperature is slowly varying 
across the main sequence, as long as the star remains convective. 

Starting from equation (\ref{tvmr}), we can derive two additional
scaling relations. First, we can use the surface boundary condition of
equation (\ref{surfacebc}) to eliminate the dependence on the radius
$R_\ast$ in favor of the luminosity $L_\ast$. This substitution yields
the expression  
\be 
\teff \propto \mu^{5/21} M_\ast^{3/21} L_\ast^{-1/42} \,.
\label{tvml} 
\ee
Alternatively, as a consistency check, we can use the fact that the
mass-luminosity relation has the form $L_\ast\sim M_\ast^2$, which
implies that $R_\ast\sim M_\ast$. Using this latter result in equation
(\ref{tvmr}), we find 
\be 
\teff \propto \mu^{5/23} M_\ast^{2/23} \,.
\label{tvmass} 
\ee

Note that the exponents in all of the above scaling laws are small, so
that the photospheric temperature $\teff$ is nearly constant across
the range of stellar masses. The surface temperature is not exactly
constant, however, but instead slowly varies with stellar mass. The
main sequence spans four orders of magnitude in mass, but the
corresponding range in surface temperature according to equation
(\ref{tvmass}) is only a factor $f\approx10^{8/23}\approx2.2$. This
range in $\teff$ agrees with that seen in the H-R diagram (see Figure
\ref{fig:mainsequence}).  

The above derivation is approximate. More complicated treatments
\cite{kippenhahn,hansen} include the fact that the stellar photosphere
is actually a thin radiative layer that matches smoothly onto the
convective interior just below the surface. Other approximations for
the opacity are also used, but as long as the opacity is a sensitive
function of temperature, the scaling exponents are small, so that 
$\teff$ varies slowly with stellar mass. 

\section{Degenerate Nuclear Burning Stars}
\label{sec:degenuke} 

The numerical simulations show that for sufficiently large enhancement
factors, stars with low masses in the approximate range $M_\ast$ = 1
-- 10 $M_J$ can sustain nuclear fusion at highly attenuated
rates. Unlike somewhat larger stars in the range $M_\ast$ = 10 -- 100
$M_J$, which are also operational, these very low-mass objects are
partially degenerate, with pressure contributions from both the ideal
gas law and electrostatic forces, but nonetheless can generate power. 
This Appendix outlines the physics of these unusual stars.

To a first approximation, the stars have constant radius as a function 
of mass (for sufficiently small $M_\ast \le 10 M_J$). Fully degenerate 
stars would have the mass-radius relation $R_\ast\sim M_\ast^{-1/3}$. 
The combination of thermal pressure for higher masses and couloumb 
forces for lower masses conspire to produce $R_\ast(M_\ast)\approx$
{\it constant}. We can model the interior structure --- the density 
distribution --- with a function of the form 
\be
\rho(r) = \rho_c p (\xi)  \qquad {\rm where} \qquad 
\xi = {r \over R_\ast}\,,
\ee
and where $\rho_c$ is the central density. It is conveninent to
define dimensionless integrals of the form
\be
I_k \equiv \int_0^1 p^k \xi^2 d\xi\,. 
\label{intdef} 
\ee
With these definitions, the stellar mass is given by the expression 
\be
M_\ast = 4\pi R_\ast^3 \rho_c I_{1} \,, 
\ee
where $I_1$ is of order unity. The stellar luminosity arising from
nuclear reactions depends on the power generated per unit volume
$\varepsilon$, i.e., the luminosity density. For a given nuclear
reaction, this quantity can be written in the form 
\be
\varepsilon(r) = \Gamma \rho^2 \Phi^2 \exp[-\Phi] \,,
\ee
where the composite variable $\Phi$ is defined by equation
(\ref{phidef}).  For proton-proton fusion under typical stellar
conditions, the constant $\Gamma_0 \approx 2200$ cm$^5$ s$^{-3}$
g$^{-1}$. In general, with enhanced cross sections for nuclear
burning, we write the factor in the form $\Gamma = X \Gamma_0$, 
where $X$ is the enhancement factor. 

For these partially degenerate stars, radiative transport through the
stellar interior is efficient.  As a result, most of the stellar
volume attains a single temperature, denoted here as $T_c$. In the
absence of electron screening (considered below), the total luminosity
is given by the expression 
\be
L_\ast = \Gamma 4\pi R_\ast^3 \rho_c^2 \Phi^2 \exp[-\Phi] I_2 \,,
\ee
where the dimensionless integral $I_2$ is defined through 
equation (\ref{intdef}). If we combine the above results, 
this baseline luminosity can be written in the form 
\be
L_\ast = X \Gamma_0 {M_\ast^2 \over 4\pi R_\ast^3} 
{I_2 \over I_1^2} \Phi^2 \exp[-\Phi] \,, 
\ee
which can be evaluated to obtain 
\be
L_\ast \approx (3L_\odot) X \left({M_\ast \over M_J}\right)^2 
\left({R_\ast \over R_J}\right)^{-3} \Phi^2 \exp[-\Phi]\,.
\label{vlmlum} 
\ee 

At the high density and relatively low temperatures of degenerate
stellar interiors, electron screening can be important. Following
classic treatments \cite{dewitt,graboske}, we define a screening
parameter
\be
\Lambda = \left(4\pi n_e\right)^{1/2} 
\left({e^2\over kT}\right)^{3/2} \,. 
\label{lamdef} 
\ee
Because of the screening effect, the Coulomb suppression of nuclear 
reactions is smaller, so that the reactions rates are enhanced. 
The enhancement factor can be written in the form 
\be
f = \exp[h(\Lambda)] \,,
\ee
where calculation of the function $h$ is complicated. In the regime of
strong screening, which is applicable to the degenerate low-mass stars
of interest, the screening funciton can be fit with the simple form
\be
h = 0.836 \Lambda^{2/3} - 0.19 \,. 
\ee
The enhancement factor competes with the Coulomb repulsion term 
and allows nuclear reactions to take place at lower temperatures. 
The combination of these two effects takes the form 
\be
\Phi^2 \exp[-\Phi] f = \Phi^2 \exp[-\Phi + h(\Lambda)] 
\qquad \qquad \qquad \qquad \qquad \qquad \qquad \qquad 
\ee
$$
= 7.44 \left( {E_G\over4kT} \right)^{2/3} 
\exp\left[ 0.836 \left(4\pi n_e\right)^{1/3} 
\left({e^2\over kT}\right) - 
3 \left( {E_G\over4kT} \right)^{1/3} \right] \,. 
$$

The star reaches a long-term steady-state configuration when the 
luminosity generated by nuclear fusion can compensate for the energy
lost by cooling. As noted previously, these objects have a nearly
degenerate configuration with efficient energy transport, so that most
of the stellar volume reaches a single temperature $T_c$.  The
luminosity due to cooling in a degenerate stellar object can be
written in the from 
\be
L_{cool} \sim M^a T_c^b \,, 
\ee
where the power-law indices $(a,b)$ have been estimated to be (1,7/2)
for white dwarfs \citep{hansen} and (6/5,9/2) for brown dwarfs
\citep{auddy}. For definiteness, we consider indices (1,4) in the
following discussion.

If we set the cooling rate (cooling luminosity) equal to the
luminosity generated by nuclear reactions, the required value of
$\Phi$ and hence the operating temperature of the star are determined.
The value is the given by the solution to an equation of the form 
\be
(M_\ast/M_J) X \Phi^{14} \exp[-\Phi+h(\Lambda)] = C \, .
\ee
The required value of $\Phi$ thus depends only logarithmically on the
stellar mass $M_\ast$ and the nuclear enhancement factor $X$. As a
result, the central temperature $T_c$ varies slowly with mass.  If
ones uses the screening factor derived above in conjunction with the
luminosity given by equation (\ref{vlmlum}), the central temperature
$T_c\sim10^5$ K leads to the stellar luminosities indicated by the
\mesa~ simulations for enhancement factor $X=10^{15}$. This central
temperature is about an order of magnitude lower than that associated
with thermonuclear fusion (e.g, deuterium burning at $T_c\approx10^6$
K) and is roughly consistent with the stellar evolution simulations
(see Figure \ref{fig:tempvtime}). Nonetheless, we lack a definitive 
calculation of the cooling rates, and the luminosity is exponentially 
sensitive to the temperature $T_c$, so this agreement is approximate. 

Finally, since these nearly degenerate stars have nearly constant
radius, the form of the lower main sequence in the H-R diagram can be
readily understood. Using $R_\ast$ = {\it constant} in conjunction
with the outer boundary condition from equation (\ref{surfacebc}), the
stellar luminosity and surface temperature obey the relation 
$L_\ast \sim T_\ast^4$. This expresion agrees with the slope of the
lower main sequence shown in the H-R diagram in Figure
\ref{fig:mainsequence}.

\section{Stellar Lifetimes versus Enhancement Factor} 
\label{sec:timevx} 

The stellar evolution simulations in the main text indicate that
enhanced cross sections for nuclear reactions result in shorter
stellar lifetimes for a given stellar mass, but the range of allowed
stellar masses becomes larger. Since the stars with the lowest mass
live the longest, these two effects compete. This Appendix considers
the balance between these two behaviors. In the limiting case where
all stars have access to one and only one nuclear burning reaction,
the extension of the mass range dominates, and the stellar lifetime
would increase (slowly) with enhancement factor. In more realistic
scenarios, however, the smallest stars cannot burn $^3$He (into
$^4$He) so that their lifetimes are somewhat shorter. As a result, the
lifetime of the longest-lived star is always of order trillions of
years, nearly independent of the enhancement factor, for the ranges of
$X$ relevant to this paper.

In order to show this dependence on the enhancement factor, we can use
the semi-analytic stellar structure model developed earlier
\cite{adams2008,adams2016,adamsreview}. The minimum stellar mass 
can be written in the form 
\be
M_{\ast{\rm min}} = 6 (3\pi)^{1/2} \left({4\over5}\right)^{3/4} 
\left({kT_{\rm nuc} \over m_e c^2}\right)^{3/4} 
\alpha_G^{-3/2} \mpro \,, 
\label{massminone} 
\ee
where $T_{\rm nuc}$ is the temperature required for nuclear reactions
to take place in the stellar core. In the context of this
semi-analytic model, the nuclear burning temperature of the star with
the minimum mass is given by the solution to an equation of the form 
\be
\Phi I(\Phi) = {A \over X} \,,
\label{phisol} 
\ee
where the quantity $\Phi$ determines the temperature as defined 
through equation (\ref{phidef}), $X$ is the enhancement factor, 
and $I(\Phi)$ is an integral function of the variable $\Phi$ 
\cite{adams2008,adams2016}. For our universe, the constant $A$ 
is given (approximately) by 
\be
A = N_T \left( {\hbar^3\over c^2} \right) 
\left( {1 \over \mpro m_e^3} \right) 
\left( {G \over \kappa_0 \conlum} \right) \,,
\ee
where $N_T$ is a dimensionless constant, $\kappa_0$ is a fiducial 
value of the stellar opacity, and $\conlum$ sets the nuclear
reaction rate. Using the same stellar model, the stellar lifetime 
has the form 
\be
t_\ast = N_\ast \hbar^3 c^4 \kappa_0 \Phi M_\ast^{-2} 
\left( G \mpro \right)^{-4} \propto \Phi^{11/2} \,, 
\label{startime} 
\ee
where $N_\ast$ is another dimensionless constant. As the enhancement
factor $X$ increases, the required value of $\Phi$ from equation
(\ref{phisol}) increases, which corresponds to decreasing values of
the nuclear burning temperature. As a result, the stellar lifetime of
the minimum mass star is predicted to increase as the enhancement
factor $X$ grows larger. 

However, the dependence of the lifetime $t_\ast(X)$ on the enhancement
factor is logarithmic: The integral function in equation
(\ref{phisol}) can be fit with the approximate form so that the
expression becomes 
\be
\Phi^{2.3} \exp[-\Phi] = {A_f \over X} \,.
\ee
To leading order, we find that $\Phi\sim\Phi_0+\ln{X}$, where
$\Phi_0\approx38$ is the value for the smallest stars in our universe
(see \cite{phillips} and Section \ref{sec:dicross}). As a result, the
nuclear burning temperature is a slowly decreasing function of the
enhancement factor $X$, and the stellar lifetime scales according 
to $t_\ast\sim(\Phi_0+\ln{X})^{11/2}$.

It is important to keep in mind that the semi-analytic model used here
does not take into account several complications. First, it assumes
only a single nuclear burning species. As found in this paper,
however, the lowest mass stars, with large enhancement factors $X$,
can burn their protons into $^3$He, but are not massive enough for the
core to produce $^4$He. The lower energy yield (for $^3$He versus
$^4$He) leads to a lower efficiency of energy conversion and a
somewhat shorter lifetime.  This correction largely compensates for
the logarithmic increase, thereby leading to maximum stellar lifetimes
that are roughly constant with $X$. Significantly, however, these
maximum stellar lifetimes are still measured in trillions of years.

\end{document}